\documentclass[preprint,prd,tightenlines,nofootinbib,superscriptaddress]{revtex4}

\usepackage{color}
\definecolor{red}{rgb}{1,0,0}

\usepackage{amsmath,bm}

\begin{document}


\preprint{\vbox{
\hbox{CERN-PH-TH/2007-014}
}  }

\title{Parton distribution function for quarks in an $s$-channel approach}
\author{F.\ Hautmann}
\affiliation{CERN, Physics Department, TH Division,  
CH-1211 Geneva 23, Switzerland}
\affiliation{Institut f{\" u}r Theoretische Physik, 
Universit{\" a}t Regensburg, D-93040 Germany}
\author{Davison E.\ Soper}
\affiliation{Institute of Theoretical Science,
University of Oregon, Eugene OR 97403, USA}

\begin{abstract}
We use an $s$-channel picture of hard hadronic collisions  
to  investigate the parton distribution function for quarks at small momentum
fraction $x$, which corresponds to very high energy scattering. We study the
 renormalized quark distribution at one loop in this approach. 
 In the high-energy picture, the quark distribution function is 
 expressed in terms of a Wilson-line correlator that represents 
 the cross section for a color dipole to scatter from the proton. We 
 model this Wilson-line correlator in a saturation model. We relate 
 this representation of the quark distribution function to the 
 corresponding representation of the structure function $F_T(x,Q^2)$ 
 for deeply inelastic scattering.    
\end{abstract}

\pacs{}

\maketitle

\section{Introduction}
\label{sec:intro}

Proton structure at large $Q^2$ and 
small Bjorken $x$ has been extensively investigated 
in experiments at HERA. This program is of great intrinsic 
interest and   provides valuable information 
for the LHC program, where the 
short-distance structure of protons and nuclei will be probed 
at TeV energies.   
Two physical pictures that seem very different from each other 
 are  used to analyze hadronic structure functions  
 for large $Q^2$ and small $x$.

There is a parton picture, in which 
the hadron  consists of partons and the partons 
undergo a hard collision that produces the final 
state.  This is reviewed in \cite{jjj}.  
This applies at large $Q^2$. The 
corresponding theoretical method is that 
of factorization. The cross section is written 
as a convolution of parton distribution 
functions, $f_{a/A}(x,Q^2)$, and a hard 
cross section for partonic scattering. 
The parton distribution functions are evaluated at 
small $x$, but the $x$ dependence is not predicted 
except insofar as it results from evolution starting 
from $f_{a/A}(x,Q_0^2)$ at a smaller virtuality scale $Q_0^2$.

There is an $s$-channel picture, in which one thinks 
of the event in the rest frame of one of the hadrons. 
See  \cite{ericemue} for recent accounts. This applies 
at small $x$. The hard 
interaction takes place far outside the hadron and 
the products of the interaction travel toward the 
hadron and interact with it. In the simplest case, 
there are effectively two objects that collide with 
the hadron. These objects carry opposite color, so 
that they can be said to constitute a ``color dipole'',  
described by a  correlator of two eikonal Wilson lines. 
An important concept here is that the cross section for the 
color dipole to scatter from the hadron can be simple 
when the transverse separation between the elements 
of the dipole are large. Then the dipole always 
scatters as long as its impact parameter is within 
the hadron radius. One speaks of the cross section 
saturating -- that is, being as large as it possibly 
could be \cite{golecrev,mue99}.

These pictures seem quite different, 
as they look at the collision in different 
reference frames, and lead to different theoretical methods, 
but they are not at all incompatible. In the region where 
their domains of validity overlap, they must describe 
the same physics. 
The aim of this paper is to connect the two pictures. 

We examine  one of the main ingredients 
used in the parton picture, namely the distribution function
for finding a quark in a hadron,  defined 
as a hadronic matrix element of a certain product of 
operators \cite{cs82}. For very small $x$, the parton system created by
this operator   is far
outside the hadron. We analyze the evolution of the system 
 using $s$-channel  methods. 
We find that the  quark distribution can be expressed as 
a Wilson-line  correlator convoluted with  
a simple  lightcone wave function. 
Moreover, we find  that this answer allows one to  
 relate with precision the seemingly dissimilar 
results for structure functions in the 
parton framework and the $s$-channel framework.

Part of the results of this analysis have been used 
in \cite{plb06} to investigate  the power  corrections 
to  structure functions  that arise from the $s$-channel  picture.  

The content of the paper is   as follows. We begin by applying 
 the  hamiltonian method~\cite{hks} to  the quark distribution function. 
This allows us to write the quark distribution  as a convolution 
of a lightcone wave function and a  matrix 
element of eikonal-line operators (Sec.~\ref{sec:qua}).   
We work in the lowest-order approximation, i.e., the dipole 
approximation. 
The convolution formula  provides  a simple  
interpretation in coordinate space  for the 
physical  process that probes 
the  distribution. The  parton distribution is 
defined by  matrix elements of operator products 
that require renormalization. We perform the renormalization 
at one loop using the $\overline{\rm MS}$ subtraction scheme 
for the ultraviolet divergences. 

The eikonal-operator matrix element 
receives contribution from both 
short distances and long distances.  We first 
analyze it   by an expansion 
in powers of $g A$ (with $g$  the strong coupling and 
$A$  the gauge field), valid at short distances 
(Sec.~\ref{sec:dipoleandglu}). This 
expansion  is useful   to carry out the matching with 
renormalization-group evolution equations.  
In particular,  
it allows us  to relate the eikonal 
 matrix element at short distances 
to a well-prescribed  integral of the 
gluon distribution function. 

Next,  we motivate and  discuss a widely used 
 approach for modeling   the  eikonal-operator  matrix 
element at large distances 
(Sec.~\ref{sec:Xi}), based on parton 
saturation~\cite{golecrev,mue99}. 
As the saturation scale in the quark sector is likely to be 
 at much  lower momenta than in the gluon sector
  (see e.g. \cite{hs00}), we critically examine 
   the validity  of the 
treatment      for 
the quark distribution,  and 
the potential breakdown of the dipole approximation 
 (Sec.~\ref{sec:crit}).

We finally discuss 
the relation of our results for the quark distribution 
 with known dipole results for structure functions 
 (Sec.~\ref{sec:strfun}). This discussion also  illustrates 
how standard factorization properties are reobtained from the 
$s$-channel point of view.

Some supplementary material is left to the appendices. In 
Appendix \ref{app:little_u} we collect calculational details 
on integrals of  lightcone wave functions. In 
Appendix~\ref{sec:algeb} we give a relation between  
products  of eikonal operators 
for color-octet and color-triplet dipoles. 
In Appendix~\ref{app:FT} we report   details on  
 applying the  hamiltonian method 
 to the hadronic matrix element of two currents.

\section{Quark distribution in the s-channel picture}
\label{sec:qua}

We study the quark distribution using 
the $s$-channel picture in the style of~\cite{hks}. 
 We start with the definition \cite{cs82} of the 
 quark distribution as a proton matrix element of a certain operator,
\begin{eqnarray}
\label{eq:fqdef}
f_{q/p}(x,\mu) &=& 
{ 1 \over 4 \pi} \left({1\over 2}\sum_s\right)
\int\! d y^- e^{i xP^+y^-}
\nonumber\\
&& \times
\langle P,s|
\bar\psi(0) Q(0) \gamma^+ Q^\dagger(y^-)\psi(0,y^-,{\bm 0})
|P,s\rangle_c .
\end{eqnarray}
Here 
for any four-vector $z^\mu$ we use  lightcone components $z^\pm$ defined as 
\begin{equation}
\label{pmkin}
z^\pm = {{z^0 \pm z^3} \over \sqrt{2} }   \; .    
\end{equation}
The proton momentum is 
\begin{equation}
\label{protmom}
P =     \left( P^+ , P^- ,  P_\perp \right)  
=   \left( P^+, \frac{M_p^2}{2P^+}, {\bm 0} \right)  \; .     
\end{equation}
 The operator $Q^\dagger$ 
is the path-ordered exponential of the color potential 
\begin{equation}
Q^\dagger(y^-) =
{\cal P}\exp\left\{
-ig\int_{y^-}^{+\infty}dz^- {A}^+_a(0,z^-,{\bm 0})t_a
\right\} \; ,
\end{equation}
where the path ordering instruction ${\cal P}$ puts fields and color
matrices with the most positive values of $z^-$ to the left. Equivalently,
following the notation of \cite{hks}, we can think of $Q^\dagger(y^-)$ 
as creating an eikonal particle that moves in the minus direction, starting  
at minus coordinate $y^-$. An eikonal particle is an imaginary particle that 
retains its plus and transverse positions no matter how much momentum
it absorbs. The
subscript $c$ on the matrix element in Eq.~(\ref{eq:fqdef}) 
 indicates that we are to take
the connected parts of the graphs, in which some partons from the proton
states communicate with the indicated operators. The operator product 
in Eq.~(\ref{eq:fqdef}) is ultraviolet divergent and requires renormalization. 
We will use the standard $\overline{\rm MS}$ prescription. The required 
subtraction at the one loop level is analyzed in Sec.~\ref{sec:renormalization}.

\subsection{The quark distribution as a forward scattering amplitude}
\label{sec:forward}

We begin by rewriting the matrix element in Eq.~(\ref{eq:fqdef}) so that it has the form of the real part of a forward scattering amplitude. To do this, we write $f_{q/p}(x,\mu)$ in two pieces,
\begin{equation}
f_{q/p}(x,\mu) = f^+_{q/p}(x,\mu) + f^-_{q/p}(x,\mu), 
\end{equation}
where
\begin{eqnarray}
f^+_{q/p}(x,\mu) &=& 
{ 1 \over 4 \pi} \left({1\over 2}\sum_s\right)
\int_0^\infty\! d y^- e^{i xP^+y^-}
\nonumber\\
&& \times
\langle P,s|
\bar\psi(0) Q(0) \gamma^+ Q^\dagger(y^-)\psi(0,y^-,{\bm 0})
|P,s\rangle_c
\end{eqnarray}
and
\begin{eqnarray}
f^-_{q/p}(x,\mu) &=& 
{ 1 \over 4 \pi} \left({1\over 2}\sum_s\right)
\int^0_{-\infty}\! d y^- e^{i xP^+y^-}
\nonumber\\
&& \times
\langle P,s|
\bar\psi(0) Q(0) \gamma^+ Q^\dagger(y^-)\psi(0,y^-,{\bm 0})
|P,s\rangle_c .
\end{eqnarray}
We note that
\begin{equation}
f^+_{q/p}(x,\mu) = [f^-_{q/p}(x,\mu)]^*.
\end{equation}
Thus $f$ is twice the real part of $f^-$:
\begin{eqnarray}
f_{q/p}(x,\mu) &=& 
{\rm Re}\, { 1 \over 2 \pi}\left({1\over 2}\sum_s\right)
\int_{-\infty}^0\! d y^- e^{i xP^+y^-}
\nonumber\\
&& \times
\langle P,s|T\!\left\{
\bar\psi(0) Q(0) \gamma^+ Q^\dagger(y^-)\psi(0,y^-,{\bm 0})
\right\}
|P,s\rangle_c  . 
\end{eqnarray}
The operator product in $f^-$ is time ordered since $y^- < 0$. 
The $T$ here indicates this time ordering. For our purposes, it is helpful to insert a factor $x$ and another $y^-$ integral:
\begin{eqnarray}
\label{eq:fqasforwardscattering}
\lefteqn{
xf_{q/p}(x,\mu) =
} 
\nonumber \\
&&
{\rm Re}\,
{ 2xP^+ \over 2 \pi} \left({1\over 2}\sum_s\right)
\int\! d y_2^- d y_1^-\ \theta(y_2^- > y_1^-)
e^{-i xP^+ (y_2^- - y_1^-)}
\int\frac{d P^{\prime +}}{(2\pi) 2 P^{\prime +}}
\nonumber\\
&& \times
\langle P',s|T\!\left\{
\bar\psi(0,y_2^-,{\bm 0}) Q(y_2^-) \gamma^+
Q^\dagger(y_1^-)\psi(0,y_1^-,{\bm 0}) \right\}|P,s\rangle_c .
\end{eqnarray}
The integral over $P^{\prime+}$ can be thought of as setting the proton
state to position $y^-  = 0$. We have thus rewritten the
original $f$, which was analogous to a total cross section, as a Green
function analogous to a forward scattering amplitude. Our next task is to break the scattering amplitude into parts that can be analyzed separately.

\subsection{Decomposition of the gluon field}
\label{sec:gluondecomposition}

The Fourier transformed operator 
$Q^\dagger(y_1^-)\psi(0,y_1^-,{\bm 0})$ 
in Eq.~(\ref{eq:fqasforwardscattering}) 
creates an antiquark and an eikonal particle with 
a total plus-momentum $xP^+$. We consider that $x$ is very small, say $10^{-3}$. That means that the typical distance $y_1^-$ from the proton to where the antiquark and the eikonal particle are created is large, of order $1/(xP^+)$. This is way outside the proton. The antiquark and eikonal particle develop into a shower of partons with minus-momenta of order $k^- = (k_\perp^2 + k^2)/(2k^+) \sim m^2/(x P^+)$, where $k_\perp$ is the transverse momentum of the parton and $k^2$ is its virtuality and we take both of these to be of order $m^2 \equiv (300\ {\rm MeV})^2$. Thus the partons created by the original operator have very large minus-momenta. We will speak of them as ``fast'' partons. As noted, the fast partons travel a long distance in $y^-$ before meeting the proton.

\begin{figure}[htb]
\vspace{105mm}
\includegraphics{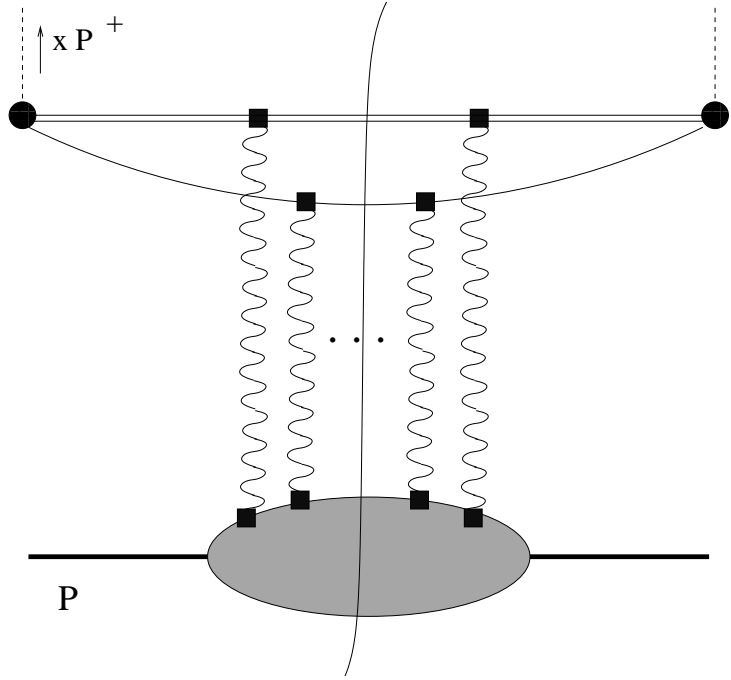}
\caption{Quark distribution in the $s$-channel picture. 
The black circles   at the top represent the insertion of  
the operators in Eq.~(\ref{eq:fqasforwardscattering}).   
The double straight line represents the eikonal. 
Any number of slow gluons couple the fast parton and 
the eikonal  to the proton.}
\label{fig:sec2}
\end{figure}

When the fast partons meet the proton, 
they scatter from the gluon field of the proton, 
as depicted in Fig.~\ref{fig:sec2}. The 
gluon field of the proton consists of ``slow'' gluons, 
with plus-momenta much larger than $xP^+$. (Then the 
minus-momenta of these gluons, $k^- = (k_\perp^2 + k^2)/(2k^+)$ is much 
smaller than $m^2/(x P^+)$, assuming again that $k_\perp^2$ and $k^2$ are 
of order $m^2$.) We represent the gluon field produced by the proton as 
an external field ${\cal A}^\mu(x)$ and consider the quantity
\begin{eqnarray}
\label{UAdef}
U[{\cal A}] &=&
{2 xP^+ \over 2 \pi}
\int\! d y_2^- d y_1^- e^{-i xP^+ (y_2^- - y_1^-)}
\nonumber\\
&& \times
\left\{
\langle 0|
\bar\psi(0,y_2^-,{\bm 0}) Q(y_2^-) \gamma^+
Q^\dagger(y_1^-)\psi(0,y_1^-,{\bm 0}) |0\rangle _{\cal A} \right.
\nonumber\\
&&\quad  -
\left.
\langle 0|
\bar\psi(0,y_2^-,{\bm 0}) Q(y_2^-) \gamma^+
Q^\dagger(y_1^-)\psi(0,y_1^-,{\bm 0}) |0\rangle _{0} \right\}.
\end{eqnarray}
This is the amplitude for the fast partons to be created by the operator $Q^\dagger\psi$, scatter from the external field ${\cal A}$, then be annihilated by the conjugate operator $\bar \psi Q$. In the second term, we subtract a no-scattering term with the external field set to zero, in accordance with the instruction to take only connected graphs. Then the quark distribution is a proton matrix element of $U[A]$, with the external field ${\cal A}$ replaced by the quantum field $A$,
\begin{equation}
xf_{q/p}(x,\mu) =
{\rm Re}\,
\left({1\over 2}\sum_s\right)
\int\frac{d P^{\prime +}}{(2\pi) 2 P^{\prime +}}
\langle P',s|
U[A]
|P,s\rangle .
\label{fieldinproton}
\end{equation}

There is, of course, a catch in this. There is only one gluon field $A^\mu(x)$. We need to divide it into two pieces, one associated with the fast partons and one associated with the proton. To do this, we choose a momentum fraction $x_c$. Gluons with plus-momenta smaller than $x_c P^+$ are associated with the fast partons. Gluons with plus-momenta larger than $x_c P^+$ are associated with the proton and included in the external field ${\cal A}$ in Eq.~(\ref{UAdef}). In order for the approximations discussed below to work, we need $x \ll x_c$. It is perhaps easiest to think about the physics taking $x_c$ to be of order $1/(R_p P^+)$, so that the proton's field is considered to have a spatial extent of the order of the proton radius, $R_p$. However, in the end we will want to take $x_c \ll 1$ and in fact let $x_c$ be pretty close to $x$.

We will, in fact, not need to be very specific about how to implement the division at momentum fraction $x_c$. For our discussion of the quark distribution, working at lowest order in perturbation theory, we are saved from sensitivity to the splitting method by the fact that the $g \to q\bar q$ Altarelli-Parisi splitting function does not have a soft singularity. If we worked with the gluon distribution or with the quark distribution to higher order, we would need a more sophisticated analysis.

\subsection{The evolution operator $U$ at high energy}
\label{sec:evolU}

The function $U[{\cal A}]$  
can be written in the interaction picture with
${\cal A}$ as the perturbation:
\begin{eqnarray}
U[{\cal A}] &=&
{ 2xP^+ \over 2 \pi}
\int\! d y_2^- d y_1^- \theta(y_2^- - y_1^-)\,
e^{-i xP^+ (y_2^- - y_1^-)}
\nonumber\\
&& \times
\langle 0| U(\infty,y_2^-)
\bar\psi(0,y_2^-,{\bm 0}) Q(y_2^-) \gamma^+ 
\nonumber\\
&& \times\ \ \ \ 
U(y_2^-,y_1^-)
Q^\dagger(y_1^-)\psi(0,y_1^-,{\bm 0})
U(y_1^-,\infty)|0\rangle 
\nonumber\\
&& - ({\rm same\ with\ } {\cal A} = 0).
\end{eqnarray}
This is without approximation. Now we 
recognize that for small $x$, only
$y_2^- \gg r^-$ and $y_1^- \ll - r^-$ 
are important, where $r^-$ is the
effective radius of the proton's field  in the 
longitudinal direction, 
$r^- = 1 / ( x_c P^+ )$. The external
field is concentrated in $|y^-| < r^-$, so we have
\begin{eqnarray}
U[{\cal A}] &\approx&
{ 2xP^+ \over 2 \pi}
\int_0^\infty \!d y_2^- \int_{-\infty}^0\! d y_1^- 
e^{-i xP^+ (y_2^- - y_1^-)}
\nonumber\\
&& \times
\langle 0|
\bar\psi(0,y_2^-,{\bm 0}) Q(y_2^-) \gamma^+ 
[U(\infty,-\infty) - 1]
Q^\dagger(y_1^-)\psi(0,y_1^-,{\bm 0})
|0\rangle .\ \ \ \
\end{eqnarray}
Here we have subtracted the no-scattering term as ``$-1$''.

At this stage, our quantum fields are evolving with full QCD [not
including the external field, which is represented in
$U(\infty,-\infty)$]. Let us now expand this evolution in powers of
$\alpha_s$ and take just the Born term. Then the fields evolve with just
the free field hamiltonian. We can insert intermediate states, and at
this level of approximation, the intermediate states contain just one
antiquark and the one eikonal particle ${\cal E}$. We get
\begin{eqnarray}
U[{\cal A}] &\approx&
{ 2xP^+ \over 2 \pi}
\int_0^\infty \!d y_2^- \int_{-\infty}^0\! d y_1^-
e^{-i xP^+ (y_2^- - y_1^-)}
\nonumber\\
&& \times
(2\pi)^{-6}
\int_0^\infty\! { dp_2^- \over 2p_2^-} \int\!d{\bm p}_2
\int_0^\infty\! { dp_1^- \over 2p_1^-} \int\!d{\bm p}_1
\sum_{s_1s_2}
\nonumber\\
&& \times
\langle 0|
\bar\psi(0,y_2^-,{\bm 0}) Q(y_2^-) \gamma^+ 
|p_2^-,{\bm p}_2,s_2, {\cal E}\rangle
\nonumber\\
&& \times
\langle p_2^-,{\bm p}_2,s_2, {\cal E}|
U(\infty,-\infty) - 1
|p_1^-,{\bm p}_1,s_1, {\cal E}\rangle 
\nonumber\\
&& \times
\langle p_1^-,{\bm p}_1,s_1, {\cal E}|
Q^\dagger(y_1^-)\psi(0,y_1^-,{\bm 0})
|0\rangle  .
\end{eqnarray}
Taking into account that particle 1 has plus momentum $p_1^+ = {\bm
p}_1^2/(2p_1^-)$ while particle 2 has plus momentum $p_2^+ = {\bm
p}_2^2/(2p_2^-)$,  we can evaluate the dependence of the matrix elements on
$y_1^-$ and $y_2^-$ as
\begin{eqnarray}
U[{\cal A}] &\approx&
{ 2xP^+ \over (2 \pi)^7}
\int_0^\infty\! { dp_2^- \over 2p_2^-} \int\!d{\bm p}_2
\int_0^\infty\! { dp_1^- \over 2p_1^-} \int\!d{\bm p}_1
\sum_{s_1s_2} 
\nonumber\\
&& \times
\int_0^\infty \!d y_2^- \int_{-\infty}^0\! d y_1^-
e^{-i (xP^+ + p_2^+) y_2^-}
e^{+i (xP^+ + p_1^+) y_1^-}
\nonumber\\
&& \times
\langle 0|
\bar\psi(0) Q(0) \gamma^+ 
|p_2^-,{\bm p}_2,s_2, {\cal E}\rangle
\nonumber\\
&& \times
\langle p_2^-,{\bm p}_2,s_2, {\cal E}|
U(\infty,-\infty) - 1
|p_1^-,{\bm p}_1,s_1, {\cal E}\rangle 
\nonumber\\
&& \times
\langle p_1^-,{\bm p}_1,s_1, {\cal E}|
Q^\dagger(0)\psi(0)
|0\rangle  .
\end{eqnarray}
We can now perform the $y^-$ integrations to produce energy denominators:
\begin{eqnarray}
U[{\cal A}] &\approx&
{ 2xP^+ \over (2 \pi)^7} 
\int_0^\infty\! { dp_2^- \over 2p_2^-} \int\!d{\bm p}_2
\int_0^\infty\! { dp_1^- \over 2p_1^-} \int\!d{\bm p}_1
\sum_{s_1s_2}
\nonumber\\
&& \times
{ -i \over xP^+ + p_2^+}\,{ -i \over xP^+ + p_1^+}
\nonumber\\
&& \times
\langle 0|
\bar\psi(0) Q(0) \gamma^+ 
|p_2^-,{\bm p}_2,s_2, {\cal E}\rangle
\nonumber\\
&& \times
\langle p_2^-,{\bm p}_2,s_2, {\cal E}|
U(\infty,-\infty) - 1
|p_1^-,{\bm p}_1,s_1, {\cal E}\rangle 
\nonumber\\
&& \times
\langle p_1^-,{\bm p}_1,s_1, {\cal E}|
Q^\dagger(0)\psi(0)
|0\rangle  .
\end{eqnarray}

For the factor giving the interaction of the partons with the external
field, we have
\begin{eqnarray}
\lefteqn{
\langle p_2^-,{\bm p}_2,s_2, {\cal E}|
U(\infty,-\infty) - 1
|p_1^-,{\bm p}_1,s_1, {\cal E}\rangle 
 = }
\nonumber\\ 
&&\hskip 1 cm
2\pi\,2p_1^-\,\delta(p_1^- - p_2^-)
\delta_{s_1s_2}
[\tilde F({\bm p}_1 - {\bm p}_2)^\dagger
F({\bm 0})
- (2\pi)^2 \delta({\bm p}_1 - {\bm p}_2)
],\ \ 
\end{eqnarray}
where
\begin{equation}
F({\bm \Delta}) = {\cal P}\exp\left\{
-ig\int_{-\infty}^{+\infty}dz^- {\cal A}^+_a(0,z^-,{\bm \Delta})t_a
\right\}
\end{equation}
and
\begin{equation}
\tilde F({\bm k}) =
\int d{\bm \Delta}\,e^{i{\bm k}\cdot {\bm \Delta}}F({\bm \Delta}).
\end{equation}

This gives
\begin{equation}
U[{\cal A}] \approx \int d{\bm \Delta}\
\frac{1}{N_c}\, {\rm Tr}
[1 - F({\bm \Delta})^\dagger F({\bm 0})]\
u({\bm \Delta}),
\end{equation}
where
\begin{eqnarray}
u({\bm \Delta}) &=&
{ 4xP^+ \over (2 \pi)^6} 
\int_0^\infty\! dp^-  
\int\!d{\bm p}_2\int\!d{\bm p}_1
\sum_{s}
e^{i{\bm \Delta}\cdot({\bm p}_1 - {\bm p}_2)}
\nonumber\\
&& \times
{ p^- \over (2xP^+p^- + {\bm p}_2^2 )(2xP^+p^- + {\bm p}_1^2)}
\nonumber\\
&& \times
\langle 0|
\bar\psi(0) Q(0) \gamma^+ 
|p^-,{\bm p}_2,s, {\cal E}\rangle
\langle p^-,{\bm p}_1,s, {\cal E}|
Q^\dagger(0)\psi(0)
|0\rangle  .
\end{eqnarray}
Thus
\begin{equation}
\label{eq:fisuXi}
xf_{q/p}(x,\mu) =
{\rm Re}
\int d{\bm \Delta}\
u({\bm \Delta})\
\Xi_I({\bm\Delta}),
\end{equation}
where
\begin{equation}
\label{eq:XiIdef}
\Xi_I({\bm\Delta}) = 
\left({1\over 2}\sum_s\right)
\int\frac{d P^{\prime +}}{(2\pi) 2 P^{\prime +}}
\langle P',s|
\frac{1}{N_c}\, {\rm Tr}
[1 - F({\bm \Delta})^\dagger F({\bm 0})]
|P,s\rangle .
\end{equation}
Here $F({\bm \Delta})$ is now defined with the quantum field $A$,
\begin{equation}
F({\bm \Delta}) = {\cal P}\exp\left\{
-ig\int_{-\infty}^{+\infty}dz^- A^+_a(0,z^-,{\bm \Delta})t_a
\right\}
.
\end{equation}
Eq.~(\ref{eq:fisuXi}) has a simple interpretation. First, 
 $u({\bm \Delta})$ is the square of the antiquark wave function, 
 giving the probability that the antiquark has reached a separation 
 $\bm \Delta$ from the eikonal line by the time it reaches the hadron. 
 Second, we have a probability $\Xi_I({\bm\Delta})$ for the antiquark-eikonal 
 dipole to scatter from the proton.

\subsection{The squared wave function for the antiquark}
\label{eq:uofDelta}

We now need the function $u(\Delta)$. First, we need the operator
matrix elements: 
\begin{eqnarray}
M &\equiv&
\sum_s
\langle 0|
\bar\psi(0) Q(0) \gamma^+ 
|p^-,{\bm p}_2,s, {\cal E}\rangle
\langle p^-,{\bm p}_1,s, {\cal E}|
Q^\dagger(0)\psi(0)
|0\rangle  
\nonumber \\
&=&
\sum_s
\langle 0|
\bar\psi(0)\gamma^+ 
|p^-,{\bm p}_2,s\rangle
\langle p^-,{\bm p}_1,s|
\psi(0)
|0\rangle 
.
\end{eqnarray}
There is an implicit color trace here. 
 Restoring the color indices makes it
\begin{eqnarray}
M
&=&
\sum_s
\langle 0|
\bar\psi_\alpha(0)\gamma^+ 
|p^-,{\bm p}_2,s,\beta\rangle
\langle p^-,{\bm p}_1,s,\beta|
\psi_\alpha(0)
|0\rangle 
.
\end{eqnarray}
Writing this with spinors gives
\begin{eqnarray}
M
&=& \delta_{\alpha\beta}\delta_{\beta\alpha}
\sum_s
\bar v(p^-,{\bm p}_2,s)
\gamma^+ v(p^-,{\bm p}_1,s)
.
\end{eqnarray}
Now we need to know about the spin states. 
We use null-plane helicity
states appropriate to the $x^-$ as the ``time.'' 
These have the
normalization
\begin{equation}
\bar v(p^-,{\bm p}_2,s')
\gamma^- v(p^-,{\bm p}_1,s)
= 2 p^- \delta_{ss'}  . 
\end{equation}
Thus
\begin{eqnarray}
M
&=& \delta_{\alpha\beta}\delta_{\beta\alpha}
\sum_{ss'}
\bar v(p^-,{\bm p}_2,s)
\gamma^+ v(p^-,{\bm p}_1,s' )\delta_{ss'}
\nonumber\\
&=&
\frac{N_c}{2p^-}
\sum_{ss'}
\bar v(p^-,{\bm p}_2,s)
\gamma^+ v(p^-,{\bm p}_1,s' )\
\bar v(p^-,{\bm p}_2,s')
\gamma^- v(p^-,{\bm p}_1,s)
\nonumber\\
&=&
\frac{N_c}{2p^-}{\rm Tr}\{
\rlap{/}p_1\gamma^-\rlap{/}p_2\gamma^+\}
\nonumber\\
&=&
\frac{N_c}{2p^-}{\rm Tr}\{
\rlap{/}p_{1,T}\,\gamma^-\rlap{/}p_{2,T}\,\gamma^+\}
\nonumber\\
&=&
\frac{2N_c}{p^-}\  {\bm p}_1\cdot {\bm p}_2
.
\end{eqnarray}
Thus
\begin{eqnarray}
u({\bm \Delta}) &=&
{ 2x\,2P^+ \over (2 \pi)^6} 
\int_0^\infty\! dp^-  
\int\!d{\bm p}_2\int\!d{\bm p}_1
e^{i{\bm \Delta}\cdot({\bm p}_1 - {\bm p}_2)}
\nonumber\\
&& \times
{ p^- \over (2xP^+p^- + {\bm p}_2^2 )(2xP^+p^- + {\bm p}_1^2)}
\nonumber\\
&& \times
\frac{2N_c}{p^-}\  {\bm p}_1\cdot {\bm p}_2 
\nonumber\\
&=&
{ 4N_c \over (2 \pi)^6} 
\int_0^\infty\! d\Lambda^{\!2}  
\int\!d{\bm p}_2\int\!d{\bm p}_1
e^{i{\bm \Delta}\cdot({\bm p}_1 - {\bm p}_2)}
\nonumber\\
&& \times
{ {\bm p}_1\cdot {\bm p}_2 \over (\Lambda^{\!2} + {\bm p}_2^2
)(\Lambda^{\!2} + {\bm p}_1^2)}
,
\end{eqnarray}
where we have defined $\Lambda^{\!2} = 2xP^+p^-$.
Extending this to $4-2\epsilon$ dimensions, we have
\begin{equation}
\label{eq:uintegral}
u({\bm \Delta}) =
{4N_c \mu^{4\epsilon} \over (2 \pi)^{6-4\epsilon} } 
\int_0^\infty\! d\Lambda^2  
\int\!d^{2-2\epsilon}{\bm p}_2\int\!d^{2-2\epsilon}{\bm p}_1\
e^{i{\bm \Delta}\cdot({\bm p}_1 - {\bm p}_2)}\
{ {\bm p}_1 \cdot {\bm p}_2 \over 
(\Lambda^2 + {\bm p}_2^2 )
(\Lambda^2 + {\bm p}_1^2)} .
\end{equation}
We perform the integration separately in Appendix~\ref{app:little_u}. We find
\begin{equation}
\label{eq:uresult}
u({\bm \Delta}) =
{N_c \over 3 \pi^4 }\, \frac{1}{\Delta^4} \
\left(\pi\mu^2 \Delta^2\right)^{2\epsilon}
\frac{\Gamma(2-\epsilon)^2}{1-2\epsilon/3}
.
\end{equation}

\subsection{Renormalization of the quark distribution}
\label{sec:renormalization}

Using the result (\ref{eq:uresult}) for $u(\bm \Delta)$, we have
\begin{equation}
\label{xfq}
xf_{q/p}(x,\mu) =
{   N_c \over 3 \pi^4}\
\frac{\Gamma(2-\epsilon)^2}{1-2\epsilon/3}\
\mu^{-2\epsilon}\!\int\! d^{2-2\epsilon}{\bm \Delta}\
{ 1 \over \Delta^4}\,\left(\pi\mu^2 \Delta^2\right)^{2\epsilon}
\Xi_I({\bm \Delta})  - {\rm{UV}} 
.
\end{equation}
Here there is an ultraviolet divergence from the small $\Delta$ integration region. The notation indicates that we should renormalize the divergence by subtracting a UV counter term.

The standard definition of the parton distribution functions 
gives these functions as hadron matrix elements of operator products 
that must be renormalized \cite{cs82}. It is thus not a surprise that 
we have a divergent integral in Eq.~(\ref{xfq}). The 
standard treatment is to apply  $\overline {\rm MS}$  renormalization. 
At the one loop level at which we work here, this means performing 
the integrals in $4-2\epsilon$ dimensions and subtracting a 
counter-term of the form
\begin{equation}
{\rm ``UV"} = {\it const.}\times \frac{1}{\epsilon}\
\frac{(4\pi)^\epsilon}{\Gamma(1-\epsilon)}
.
\end{equation}
In this section, we implement this subtraction, turning it (approximately) into a cutoff on $|\bm \Delta|$.

We are eliminating only the divergence from the innermost loop in the Feynman diagrams that define $xf_{q/p}(x,\mu)$, so we treat the outer loops in $\Xi_I({\bm \Delta})$ as containing only soft momenta. For this reason, we treat $\Xi_I({\bm \Delta})$ as being an analytic function of ${\bm \Delta}$ near ${\bm \Delta} = 0$. We thus write
\begin{equation}
\begin{split}
\Xi_I({\bm \Delta}) ={}& 
\Xi_I(0) + \Delta^i [\partial_i \Xi_I({\bm \Delta})]_{\Delta = 0}
+\frac{1}{2}\Delta^i\Delta^j 
[\partial_i \partial_j\Xi_I({\bm \Delta})]_{\Delta = 0}
\\&+
\frac{1}{3!}\Delta^i\Delta^j\Delta^k
[\partial_i \partial_j\partial_k\Xi_I({\bm \Delta})]_{\Delta = 0}
+ R(\Delta)
.
\end{split}
\end{equation}
(We follow the convention that indices $i,j,k$ are summed from 1 to 2 or, with dimensional regularization, from 1 to $2 - 2\epsilon$.) The remainder, $R(\Delta)$, goes to zero like $\Delta^4$ as $\Delta \to 0$. The first term vanishes because $\Xi_I(0) = 0$ by construction. The second and fourth terms vanish upon integrating over $\bm \Delta$. In the third term, under the integration over $\bm \Delta$, we can replace
\begin{equation}
\Delta^i\Delta^j \to \frac{1}{2-2\epsilon}\ \delta^{ij} \Delta^2
.
\end{equation}
Thus, in the small $\Delta$ integration region, we can replace
\begin{equation}
\Xi_I({\bm \Delta}) \to \frac{1}{4(1-\epsilon)}\ \Delta^2
[\partial^2_\perp\Xi_I({\bm \Delta})]_{\Delta = 0} + R(\bm \Delta)
.
\end{equation}

We introduce this approximation in the small $\Delta$ integration region, defined by $\Delta \mu < a$, where $a$ is a parameter of order 1 that we can adjust. Thus we write
\begin{equation}
\begin{split}
xf_{q/p}(x,\mu) ={}&
\frac{N_c}{3 \pi^4}\
\!\int\! d^2{\bm \Delta}\
\theta(\Delta^2\mu^2 > a^2)\
\frac{\Xi_I({\bm \Delta})}{\Delta^4} 
\\& +
\frac{N_c}{3 \pi^4}\
\!\int\! d^2{\bm \Delta}\
\theta(\Delta^2\mu^2 < a^2)\
\frac{R({\bm \Delta})}{\Delta^4}
\\
&+ \frac{N_c}{12 \pi^4}\
[\partial^2\Xi_I({\bm \Delta})]_{\Delta = 0}
\\ &\quad\times
\frac{\Gamma(2-\epsilon)^2}{(1-\epsilon)(1-2\epsilon/3)}\
\mu^{-2\epsilon}\!\int\! d^{2-2\epsilon}{\bm \Delta}\
\frac{\theta(\Delta^2\mu^2 < a^2)}{\Delta^2}
\,\left(\pi\mu^2 \Delta^2\right)^{2\epsilon}
\\&  - {\rm{UV}} . 
\end{split}
\end{equation}
In the first two terms, there is no ultraviolet divergence, so we set $\epsilon \to 0$. We can perform the integration in the third term to obtain
\begin{equation}
\begin{split}
xf_{q/p}(x,\mu) ={}&
\frac{N_c}{3 \pi^4}\
\!\int\! d^{2}{\bm \Delta}\
\theta(\Delta^2\mu^2 > a^2)\
\frac{\Xi_I({\bm \Delta})}{\Delta^4} 
\\& +
\frac{N_c}{3 \pi^4}\
\!\int\! d^{2}{\bm \Delta}\
\theta(\Delta^2\mu^2 < a^2)\
\frac{R({\bm \Delta})}{\Delta^4}
\\
&+ \frac{N_c}{12 \pi^3}\
[\partial^2\Xi_I({\bm \Delta})]_{\Delta = 0}\
\frac{1}{\epsilon}\,\frac{(4\pi)^\epsilon}{\Gamma(1-\epsilon)}
\\ &\quad\times
\left\{
\frac{\Gamma(2-\epsilon)^2}{(1-\epsilon)(1-2\epsilon/3)}\
(a^2/4)^{\epsilon}
-1
\right\}
.
\end{split}
\end{equation}
Here we have identified the UV subtraction term and written it as $-1$ inside the braces in the last term. The $-1$ removes the $1/\epsilon$ pole, but, in general, leaves a remainder that is finite as $\epsilon \to 0$. We set
\begin{equation}
\label{eq:aresult}
a = 2 e^{1/6 - \gamma} \approx 1.32657
,
\end{equation}
where $\gamma$ is the Euler constant, $\gamma \approx 0.577216$, that appears in $\Gamma(1 - \epsilon) = 1 + \epsilon\gamma + \cdots$. With this choice, we cancel the finite term and leave
\begin{equation}
\begin{split}
xf_{q/p}(x,\mu) ={}&
\frac{N_c}{3 \pi^4}\
\!\int\! d^{2}{\bm \Delta}\
\theta(\Delta^2\mu^2 > a^2)\
\frac{\Xi_I({\bm \Delta})}{\Delta^4} 
\\& +
\frac{N_c}{3 \pi^4}\
\!\int\! d^{2}{\bm \Delta}\
\theta(\Delta^2\mu^2 < a^2)\
\frac{R({\bm \Delta})}{\Delta^4}
.
\end{split}
\end{equation}

The term containing $R(\Delta)$ is needed to express the result 
of $\overline {\rm MS}$ renormalization if $1/\mu$ is of the order 
of the proton radius, $R_p$. However, the parton distribution 
function evaluated at such a renormalization scale is not 
really a very interesting object. For large values of $\mu$, 
the term containing $R(\Delta)$ is of order $1/(\mu^2 R_p^2)$ and 
can be neglected. Thus, as long as $1/(\mu^2 R_p^2) \ll 1$, 
we can write\footnote{The derivation assumes 
that $\Xi_I({\bm \Delta})$, and thus $R(\Delta)$, is 
defined with a fixed renormalization scale $\mu$. If 
we use Eq.~(\ref{eq:renormalizedfq}), however, we 
can set $\mu$ in $\Xi_I({\bm \Delta})$ to $a/\Delta$.}
\begin{equation}
\begin{split}
\label{eq:renormalizedfq}
xf_{q/p}(x,\mu) ={}&
\frac{N_c}{3 \pi^4}\
\!\int\! d{\bm \Delta}\
\theta(\Delta^2\mu^2 > a^2)\
\frac{\Xi_I({\bm \Delta})}{\Delta^4} 
.
\end{split}
\end{equation}
This is a remarkably simple formula. Almost everything is contained in the dipole scattering function $\Xi_I({\bm \Delta})$. In the following section, we will study $\Xi_I({\bm \Delta})$ for small $\Delta$, where perturbation theory can be used. Then in Sec.~\ref{sec:Xi}, we will introduce and motivate on physical grounds a well known model for $\Xi_I({\bm \Delta})$ for large $\Delta$.

\section{Dipole scattering and the gluon distribution}
\label{sec:dipoleandglu}

In this section, we investigate the dipole scattering function $\Xi_I({\bm \Delta})$ for small $\Delta$, where the use of perturbation theory is allowed. We will see that $\Xi_I({\bm \Delta})$ for small $\Delta$ is related to the gluon distribution. 

\subsection{The dipole scattering function at small $\Delta$}
\label{sec:XiatsmallDelta}

We begin by studying $\Xi_I({\bm \Delta})$ for small $\Delta$ and at lowest order in an expansion in powers of the strong coupling $g$. For the sake of generality, we consider using color matrices $t_c$ in a representation $r$ of SU(N) that need not be the fundamental representation that is appropriate for the quark distribution. We start by defining $\Xi_I^r({\bm\Delta})$ corresponding to the representation $r$,
\begin{equation}
\label{eq:XiIdef2}
(2\pi) 2 P^{\prime +}\delta(P^+ - P^{\prime +})
\delta_{ss'}\ d_r
\Xi_I^r({\bm\Delta}) = 
\langle P',s'|
{\rm Tr}
[1 - F_r({\bm \Delta})^\dagger F_r({\bm 0})]
|P,s\rangle 
,
\end{equation}
where
\begin{equation}
\label{eikFdef2}
F_r({\bm \Delta}) = {\cal P}\exp\left\{
-ig\int_{-\infty}^{+\infty}dz^- A^+_a(0,z^-,{\bm \Delta})t_a
\right\}
.
\end{equation}
The matrices $t_a$ here are in the color representation $r$ and
$d_r$ is the dimension of the representation, 
\begin{equation}
\label{drfund}
{\mbox{fundamental}}: \hspace*{0.5 cm} d_r = N_c  \hspace*{0.5 cm} , 
\end{equation}
\begin{equation}
\label{dradj}
  {\mbox{adjoint}}:   \hspace*{0.5 cm} d_r = N_c^2-1 \hspace*{0.5 cm} . 
\end{equation}
For the fundamental representation, averaging over spins and integrating over $P'$ gives the definition (\ref{eq:XiIdef}) of $\Xi_I({\bm\Delta})$ that we used for the quark distribution.

We are interested  in small ${\bm \Delta}$, for which an expansion in
powers of $ g A$ is justified.  If we limit ourselves to evaluating the
trace to the accuracy $(g A)^2$, we  can ignore the ${\cal P}$-products
and the non-commutativity of the fields in the exponent:  
\begin{equation}
\label{smallfie}
{\rm Tr}
[1 - F_r({\bm \Delta})^\dagger F_r({\bm 0})]
\approx
{\mbox{Tr}} \left[ 1 - \exp\left(
 i g \int^{\infty}_{- \infty} d z^-\, [ A_c^+(0,z^-,{\bm \Delta}) 
- A_c^+(0,z^-,{\bm 0})] \, t_c
\right) \right]
.
\end{equation}

The exponent has the expansion  around ${\bm \Delta} = 0$
\begin{eqnarray}
\label{delexpan}
A_c^+(0,z^-,{\bm \Delta}) 
- A_c^+(0,z^-,{\bm 0}) &=&  
\Delta^i [ \partial_i A_c^+(0,z^-,{\bm \Delta}) ]_{{\bm \Delta} = 0} 
\nonumber\\
&+& {1 \over 2} \Delta^i \Delta^j 
 [ \partial_i \partial_j   A_c^+(0,z^-,{\bm \Delta}) 
]_{{\bm \Delta} = 0}  + \cdots
.
\end{eqnarray}
Now we evaluate the trace to order $\Delta^2$ by expanding the exponential.  
The zeroth order term from the exponential expansion cancels against~1.  
The linear term  gives zero because ${\mbox{Tr}}\  t_c   = 0$.  The quadratic term in the exponential expansion receives an order-$\Delta^2$  contribution  only from the first-derivative term in the exponent (\ref{delexpan}), again because ${\mbox{Tr}}\  t_c   = 0$. Thus
\begin{eqnarray}
\label{quadrexp}
{\rm Tr}
[1 - F_r({\bm \Delta})^\dagger F_r({\bm 0})]
 &\simeq& - {1 \over 2} (i g)^2 \Delta^i \Delta^j 
{\mbox{Tr}} [ t_a t_b] 
\int^{\infty}_{- \infty} d z_1^- \int^{\infty}_{- \infty} d z_2^- 
\nonumber\\
&\times& 
[ \partial_i A_a^+ (0,z_1^-,{\bm \Delta}) ]_{{\bm \Delta} = 0} \,
[ \partial_j A_b^+ (0,z_2^-,{\bm \Delta}) ]_{{\bm \Delta} = 0}
\hspace*{0.2 cm} . 
\end{eqnarray}
The trace of two generators is 
\begin{equation}
\label{tracetatb}
{\mbox{Tr}} [ t_a t_b] = c_r \delta_{a b}
\hspace*{0.2 cm} ,  
\end{equation}
where $c_r$ depends on the representation: 
\begin{equation}
\label{crfund}
{\mbox{fundamental}}: \hspace*{0.5 cm} c_r = T_R = 1/2 
,
\end{equation}
\begin{equation}
\label{cradj}
{\mbox{adjoint}}:   \hspace*{0.5 cm} c_r = C_A = N_c 
.
\end{equation}
For matrix elements over hadron states carrying no transverse momentum 
we can replace 
\begin{equation}
\label{notrans}
\Delta^i \Delta^j \partial^i A_a^+ \partial^j A_b^+ \to 
{1 \over 2} {\bm \Delta}^2  \partial_j A_a^+ \partial_j A_b^+
. 
\end{equation}
Thus, to second order in $g A$ and second order in $\Delta$, we have 
\begin{eqnarray}
\label{avtrace}
 \langle P^\prime,s' | 
{\mbox{Tr}} \left[ 1 - F^\dagger_r ({\bm \Delta} ) 
F_r( {\bm 0} ) \right] | P ,s
\rangle 
&\approx& 
 {1 \over 4} g^2 c_r  {\bm \Delta}^2 
\int^{\infty}_{- \infty} d z_1^- \int^{\infty}_{- \infty} d z_2^- 
\\
&\times &
\langle P^\prime,s' | 
\partial_j A_a^+  (0,z_1^-,{\bm 0}) 
\partial_j A_a^+  (0,z_2^-,{\bm 0}) 
| P ,s\rangle  
.
\nonumber
\end{eqnarray}

At this point we make explicit our restriction on the field operators $A^\mu(x)$, namely that only modes with gluon momenta $|q^+|$ larger than $x_c P^+$ are included. (See Sec.~\ref{sec:gluondecomposition}.) This means that the reach in coordinate space in the integrations over $z_1^-$ and $z_2^-$ is limited to $|z_1^- - z_2^-| < 1/(x_c P^+)$. Thus we write
\begin{eqnarray}
\label{eq:XiIsmallDelta}
(2\pi) 2 P^{\prime +}\delta(P^+ - P^{\prime +})
\delta_{ss'}\,
\Xi_I^r({\bm\Delta})
&\approx& 
\frac{g^2}{4} \frac{c_r}{d_r}  {\bm \Delta}^2 
\int^{\infty}_{- \infty} d z_1^- \int^{\infty}_{- \infty} d z_2^- 
\nonumber\\
&\times &
\theta(|z_1^- - z_2^-| < 1/(x_c P^+))
\\
&\times &
\langle P^\prime,s' | 
\partial_j A_a^+  (0,z_1^-,{\bm 0}) \,
\partial_j A_a^+  (0,z_2^-,{\bm 0}) 
| P ,s\rangle  
.
\nonumber
\end{eqnarray}

\subsection{The gluon distribution function}
\label{sec:gluondist}

One might suspect that the right hand side of Eq.~(\ref{eq:XiIsmallDelta}), being quadratic in the gluon field, may have something to do with the gluon distribution function. Indeed, as we shall see later, the relation is well known. We can check this relation by referring directly to the definition of the gluon distribution \cite{cs82},
\begin{equation}
\label{opg}
f_{g/p} (x, \mu) = 
{1 \over 2\pi x P^+}  \int d y^-
e^{ix P^+ y^-}
\left(\frac{1}{2}\sum_s\right)
\langle P,s | 
\widetilde F_a^{+j} (0,0,{\bm 0}) 
\widetilde F_a^{+j} (0,y^-,{\bm 0}) | P,s \rangle 
, 
\end{equation}
with 
\begin{equation}
\label{Ftilde}
\widetilde F_a(y)^{+j}
=
E(y)_{ab}
F_b(y)^{+j}
,
\end{equation} 
\begin{equation}
\label{eikdef}
E(y) = 
{\cal P}
\exp\left(
- i g \int_{y^-}^\infty d z^-\, A_c^+(y^+,z^-,{\bm y})\, t_c
\right)
.
\end{equation}
We study this distribution in the $s$-channel picture.  At the lowest
order  in this picture the  gluons from the background field  couple to
the vertex measured by the operator (\ref{opg}). 

We use momentum conservation to insert a second integral over the minus coordinate in Eq.~(\ref{opg}), and we also use rotational invariance to eliminate the spin average: 
\begin{eqnarray}
\label{opgyy}
2 \pi \delta( P^{\prime +} - P^+) \delta_{ss'}f_{g/p} (x, \mu)
 &=&  
{1 \over 2\pi x P^+}  \int_{- \infty}^{\infty} d y_1^- 
\int_{- \infty}^{\infty} d y_2^- \
e^{ix P^+ (y_1^- - y_2^-) } 
\nonumber\\ 
& \times & 
\langle P^\prime,s' | 
\widetilde F_a^{+j} (0,y_2^-,{\bm 0}) 
\widetilde F_a^{+j} (0,y_1^-,{\bm 0}) | P,s\rangle 
.  
\end{eqnarray}
As in \cite{hks}, it is convenient to
rewrite $\widetilde F^{+j}$  as 
\begin{equation}
\label{newF}
\widetilde F_a^{+j}(y) = \partial^+ ( E(y)_{ab} A_b^j(y) ) 
- E(y)_{ab}\partial^j A_b^+(y)   
\end{equation}
and note that inside the integral in Eq.~(\ref{opg}), $\partial^+ $ gives $- i x P^+$. Thus, in the limit $x \to 0$, the first term in Eq.~(\ref{newF}) can be neglected. Additionally, to lowest order in a perturbative expansion, the eikonal operator $E(y)_{ab}$ is equivalent to the unit operator. Thus we replace
\begin{equation}
\label{bornx0}
\widetilde F_a^{+j}(y) \to - \partial^j A_a^+(y)
.  
\end{equation}
This gives
\begin{eqnarray}
\label{opgyy2}
2 \pi \delta( P^{\prime +} - P^+) \delta_{ss'}f_{g/p} (x, \mu)
 &\approx&  
{1 \over 2\pi x P^+}  \int_{- \infty}^{\infty} d y_1^- 
\int_{- \infty}^{\infty} d y_2^- \
e^{ix P^+ (y_1^- - y_2^-) } 
\nonumber\\ 
& \times & 
\langle P^\prime,s' | 
\partial_j A_a^+(y)(0,y_2^-,{\bm 0})\,
\partial_j A_a^+(y)(0,y_1^-,{\bm 0}) | P,s\rangle 
.  
\end{eqnarray}

We need one more approximation. For small $x$, the factor $\exp({ix P^+ (y_1^- - y_2^-) })$ is approximately 1. This is not exact, and fails for very large $(y_1^- - y_2^-)$. For $|y_1^- - y_2^-| > 1/(xP^+)$, the matrix element is a slowly varying function of $(y_1^- - y_2^-)$, so the oscillating factor $\exp({ix P^+ (y_1^- - y_2^-) })$ effectively cuts off the integral. Thus we approximate $\exp({ix P^+ (y_1^- - y_2^-) })$ by a theta function that restricts the integration to $|y_1^- - y_2^-| < 1/(xP^+)$. This gives
\begin{eqnarray}
\label{xf0}
2 \pi \delta( P^{\prime +} - P^+) \delta_{ss'}f_{g/p} (x, \mu)
 &\approx&  
{1 \over 2\pi x P^+}  \int_{- \infty}^{\infty} d y_1^- 
\int_{- \infty}^{\infty} d y_2^- \
\theta(|y_1^- - y_2^-| < 1/(xP^+))
\nonumber\\ 
& \times & 
\langle P^\prime,s' | 
\partial_j A_a^+(y)(0,y_2^-,{\bm 0})\,
\partial_j A_a^+(y)(0,y_1^-,{\bm 0}) | P,s\rangle 
.  
\end{eqnarray}

\subsection{Relation between $f_{g/p}$ and $\Xi$}
\label{sec:relation}

If we compare Eqs.~(\ref{eq:XiIsmallDelta}) and (\ref{xf0}), we see that
\begin{equation}
\label{eq:cfXigluon0}
\Xi_I^r({\bm\Delta}) \approx 
\frac{\pi g^2}{4} \frac{c_r}{d_r}  {\bm \Delta}^2\,
x_c f_{g/p} (x_c, \mu)
.
\end{equation}
Eq.~(\ref{eq:cfXigluon}) in particular implies that 
\begin{equation}
\label{eq:xirat}
{ { \Xi}_{I , \rm{fund.}} \over { \Xi}_{I , \rm{adj.}} }
\approx  { {(1/2) /3 } \over { 3 / 8} } 
=  {4 \over 9}  \ =  \ { C_F \over C_A}  
\end{equation}
for small $\Delta$ and small $x$ and to lowest perturbative order. There is a more general relation between ${\Xi}_{I, \rm{fund.}}$ and ${\Xi}_{I , \rm{adj.}}$, which we give in Appendix~\ref{sec:algeb}.

The case of interest to us here is that of the fundamental representation, which applies to the quark distribution function. For this case, the result is
\begin{equation}
\label{eq:cfXigluon}
\frac{\Xi_I({\bm\Delta})}{{\bm \Delta}^2} \approx 
\frac{\pi^2 \alpha_s}{2N_c}\,
x_c f_{g/p} (x_c, \mu)
.
\end{equation}

\subsection{Matching using the renormalization group}
\label{sec:matching}

In deriving Eq.~(\ref{eq:cfXigluon}), we have employed 
rather crude approximations relating to the integrations 
over the minus component of position for the gluon field. 
The main idea was that the  structure in matrix 
elements of $A^+(y)$ occurs for $y^-$  less than 
 $1 / (x_c P^+)$, 
so that limits on the integrations over $y^-$ should not much matter. 
If this were precisely the case, then the 
function $x_c f_{g/p} (x_c , \mu)$ that 
appears on the right hand side of Eq.~(\ref{eq:cfXigluon}) would be 
independent of $x_c$. In fact, $x_c f_{g/p} (x_c, \mu)$ 
grows slowly as $x_c$ decreases. Thus, we should try to make the relation (\ref{eq:cfXigluon}) more precise. 

We can use the scale dependence of $f_{q/p}(x,\mu)$ to provide a more precise matching condition. On one hand, we have the leading order renormalization group equation,
\begin{equation}
\frac{d}{d\log(\mu^2)}\,x f_{q/p}(x,\mu) = \frac{\alpha_s}{2\pi}
\int_x^1\! dz\, P_{qg}(z)\, 
\frac{x}{z}\,f_{g/p}\!\left(\frac{x}{z},\mu\right)
+\frac{\alpha_s}{2\pi}
\int_x^1\! dz\, P_{qq}(z)\, 
\frac{x}{z}\,f_{q/p}\!\left(\frac{x}{z},\mu\right)
.
\end{equation}
At small $x$, the gluon distribution dominates and the quark distribution is effectively $\alpha_s$ times the gluon distribution (as is, in fact, consistent with this equation). Thus the renormalization group equation can be approximated by
\begin{equation}
\frac{d}{d\log(\mu^2)}\,x f_{q/p}(x,\mu) = \frac{\alpha_s}{2\pi}\ T_R
\int_x^1\! dz\  \left[z^2 + \left(1-z\right)^2\right]\,
\frac{x}{z}\,f_{g/p}\!\left(\frac{x}{z},\mu\right)
.
\end{equation}
where we have inserted the specific form of $P_{qg}(x/y)$.

In our small $x$ approximations, $f_{q/p}(x,\mu)$ is given by Eq.~(\ref{eq:renormalizedfq}) (as long as $\mu \gg 1/R_p$). Differentiating this equation with respect to $\mu$ gives
\begin{equation}
\frac{d}{d\log(\mu^2)}\,x f_{q/p}(x,\mu)
\approx
\frac{N_c}{3\pi^3}\ 
\left[\frac{\Xi_I(\Delta)}{\Delta^2}
\right]_{\Delta = a/\mu}
.
\end{equation}
Comparing these equations gives
\begin{equation}
\label{eq:XismallDelta}
\frac{\Xi_I(\Delta)}{\Delta^2} = \frac{\pi^2 \alpha_s}{2N_c}\,
x G(x,a/\Delta)
,
\end{equation}
for $\Delta \ll R_p$, where
\begin{equation}
\label{eq:Gxdef}
xG(x,\mu) = \frac{3}{2}
\int_x^1\! dz\  \left[z^2 + \left(1-z\right)^2\right]\,
\frac{x}{z}\,f_{g/p}\!\left(\frac{x}{z},\mu\right)
.
\end{equation}
Note that the lower limit on the $z$ integral is just a reminder that $y f_{g/p}\!\left(y,\mu\right)$ vanishes for $y>1$. Note also that the integral of the weight function is
\begin{equation}
\frac{3}{2} \int_0^1\! dz\ \left[z^2 + \left(1-z\right)^2\right] = 1
.
\end{equation}
Thus $xG(x,\mu)$ is $y f_{g/p}\!\left(y,\mu\right)$ averaged over values of $y$
that are somewhat larger than $x$. If we consider a typical value of $z$ to be
1/2, then the typical value of $y$ at which the gluon distribution is evaluated
is $y = 2x$. If, for example, 
$x f_{g/p}\!\left(x,\mu\right) \propto x^{-0.3}$ for small $x$, then $xG(x,\mu) \approx 0.76 \times x f_{g/p}\!\left(x,\mu\right)$ for small $x$.

Eq.~(\ref{eq:XismallDelta}) is the same result as 
in Eq.~(\ref{eq:cfXigluon}), except that now 
$x_c f_{g/p}\!\left(x_c,\mu\right)$ is replaced 
by the more precise value, $x G(x,a/\Delta)$. Note that the matching 
condition suggests that $x_c$ be set to a value not much bigger 
than $x$. This is  in part  
because our perturbative calculation of $U({\cal A})$ was to zeroth order only. Had we worked to one more order in perturbation theory, we could have included the emission of a fast gluon with momentum fraction between $x$ and $x_c$. However, to the order to which we calculated, there were {\em no} interactions with fast gluons. Working to this order, the best choice is to include all possible gluons as slow gluons. This means setting $x_c$ to something close to $x$.  

\section{The hadronic matrix element}
\label{sec:Xi}

In this section, we motivate a widely used model for $\Xi_I(\bm \Delta)$ that applies at large $\Delta$. We begin by writing $\Xi_I(\bm \Delta)$ as an integral of a function $\Xi(\bm b,\bm \Delta)$ that has a direct physical interpretation.

\subsection{Scattering at fixed impact parameter}
\label{sec:Xiofb}

We can rewrite the hadron matrix element in Eq.~(\ref{eq:XiIdef}) by introducing an integration over an impact parameter $\bm b$. Denoting an eigenstate of transverse position by a subscript $x$, we have 
\begin{eqnarray}
\lefteqn{
\langle  P^{\prime+},{\bm 0},s|
\frac{1}{N_c}\, {\rm Tr}[1 - F({\bm \Delta})^\dagger F({\bm 0})]
| P^+,{\bm 0},s\rangle }
\nonumber\\
&=&
\langle  P^{\prime+},{\bm 0},s|
\frac{1}{N_c}\, {\rm Tr}[1 - F({\bm \Delta}/2)^\dagger F(-{\bm \Delta}/2)]
| P^+,{\bm 0},s\rangle
\nonumber\\
&=&
\frac{1}{N_c}
\int\! d{\bm b}\
{}_x\!\langle P^{\prime +},-{\bm b},s|
{\rm Tr}[1 - F({\bm \Delta}/2)^\dagger F(-{\bm \Delta}/2)]
|P^+,{\bm 0},s\rangle
\nonumber\\
&=&
{ 1 \over N_c}
\int\! d{\bm b}\
{}_x\!\langle  P^{\prime+},{\bm 0},s|
{\rm Tr}[1 - F({\bm b}+{\bm \Delta}/2)^\dagger F({\bm b}-{\bm \Delta}/2)]
|P^+,{\bm 0},s\rangle
.
\end{eqnarray}
Thus
\begin{equation}
\label{eq:Xiintegral}
\Xi_I({\bm \Delta}) = \int\! d{\bm b}\
\Xi({\bm b},{\bm \Delta})
,
\end{equation}
where\footnote{Here $\Xi({\bm b},{\bm \Delta})$ is 
the spin average (with $s' = s$) of what is called $\Xi$ in \cite{hs00}.}
\begin{equation}
\label{eq:XibDeltadef}
\Xi({\bm b},{\bm \Delta}) =
\frac{1}{N_c}\left(\frac12\sum_s\right) 
\int\frac{d P^{\prime +}}{(2\pi) 2 P^{\prime +}}\
{}_x\!\langle  P^{\prime+},{\bm 0},s|
{\rm Tr}[1 - F({\bm b}+{\bm \Delta}/2)^\dagger F({\bm b}-{\bm \Delta}/2)]
|P^+,{\bm 0},s\rangle
.
\end{equation}
The quantity $\Xi({\bm b},{\bm \Delta})$ is more suitable than $\Xi_I({\bm \Delta})$ as a quantity to model since the physics of the dipole-proton interaction should depend on $\bm b$. Given a model for $\Xi({\bm b},{\bm \Delta})$, one obtains $\Xi_I({\bm \Delta})$ by integrating over $\bm b$.

Given that $\Xi_I({\bm \Delta})$ has the behavior given by Eq.~(\ref{eq:XismallDelta}) at small $\Delta$, we can write for $\Xi({\bm b},{\bm \Delta})$ at small $\Delta$,
\begin{equation}
\label{eq:XiofBsmallDelta}
\Xi(\bm b, \bm\Delta) = \Delta^2\frac{\pi^2 \alpha_s}{2N_c}\,
x G(x,a/\Delta)\,\phi(\bm b)
,
\end{equation}
where
\begin{equation}
\int d\bm b \ \phi(\bm b) = 1
.
\end{equation}
Given that $x G(x,a/\Delta)$ is the number of gluons per unit $d\log x$ (averaged over momentum fractions somewhat larger than $x$), we interpret $x G(x,a/\Delta)\,\phi(\bm b)$ as the number of gluons per unit area $d\bm b$ and per unit $d\log x$ at a distance $\bm b$ from the center of the proton. Consistently with this interpretation, we assume that
\begin{equation}
\phi(\bm b) \ge 0
\end{equation}
and
\begin{equation}
\phi(\bm b) = 0 \quad\quad \mbox{for}\ |\bm b| > R_p.
\end{equation}
We will need a model for $\phi(\bm b)$.

\subsection{Interpretation and properties of $\Xi(\bm b, \bm\Delta)$}
\label{sec:Xiproperties}

Let us write $\Xi(\bm b, \bm\Delta)$ as
\begin{equation}
\Xi(\bm b, \bm\Delta) = 1 - T(\bm b, \bm\Delta)
.
\end{equation}
Here the 1 comes from the 1 in Eq.~(\ref{eq:XibDeltadef}). 
Then $T$ comes from the matrix element of $F^\dagger F$. 
In the language of classical optics, $T(\bm b, \bm\Delta)$ 
is the transmission coefficient for a dipole of size $\Delta$ 
impinging on the proton at impact parameter $\bm b$. 
According to the definition (\ref{eq:XibDeltadef}), 
the dipole is counted as transmitted only if the proton 
is left intact after the dipole moves through it. (This 
is the consequence of our having switched from a description of $f_{q/p}(x , \mu)$ 
as a total cross section to a description in the form of a forward scattering amplitude.) 
Based on this interpretation and on what we have already learned about $\Xi(\bm b, \bm\Delta)$, 
we expect $\Xi(\bm b, \bm\Delta)$ to have the following properties.

\begin{enumerate}

\item $T(\bm b, \bm\Delta) = 1$ for $|\bm b| > R_p + \Delta/2$.

\item $T(\bm b, \bm\Delta) = 1$ for $\Delta = 0$.

\item $T(\bm b, \bm\Delta) \approx 0$ for $|\bm b| < R_p$ with $|\bm b|$ not close to $R_p$ and $\Delta$ not small.

\item $T(\bm b, \bm\Delta) = 1 - \Delta^2\,[{\pi^2 \alpha_s}/{(2N_c)}]\,
x G(x,a/\Delta)\,\phi(\bm b) + {\cal O}(\Delta^4)$ for $\Delta \to 0$.

\end{enumerate}
Property 1 simply says that a dipole that entirely misses the proton does not interact with it and is thus perfectly transmitted. Property 2 holds because a dipole with zero separation does not have any interaction with the proton. This is the property of color transparency. Property 3 applies because a big dipole has strong interactions, so that we expect that after such a dipole moves through the proton the proton is almost never left intact. Property 4 is consistent with $T$ being 1 for $\Delta = 0$ and reflects our previously obtained perturbative result for $\Xi$ at small $\Delta$.  

\subsection{Model for $\Xi(\bm b, \bm\Delta)$}
\label{sec:satur}

There is a simple model for $T(\bm b, \bm\Delta)$ that is consistent 
with the properties listed in the previous subsection,
\begin{equation}
\label{eq:Tmodel}
T(\bm b, \bm\Delta) = \exp\left(
-\Delta^2\,\frac{\pi^2 \alpha_s}{2N_c} \,
x G(x,a/\Delta)\,\phi(\bm b)
\right)
.
\end{equation}
This is a small variation on the widely 
used {\em saturation model}~\cite{mue99,golec}, with 
 the gluon distribution   treated according to  the matching 
 of Sec.~\ref{sec:matching}.  
The same model for $\Xi(\bm b, \bm\Delta)$ is
\begin{equation}
\label{eq:ximodel}
\Xi(\bm b, \bm\Delta) = 1 - e^{ - {\bm \Delta}^2 Q_s^2 ( {\bm b} ) / 4}
,     
\end{equation}
where $Q_s(\bm b)$, known as the saturation scale, is
\begin{equation}
\label{eq:Qssq}
Q_s^2 ( {\bm b} ) = 
\frac{2\pi^2 \alpha_s}{N_c} \,
x G(x,a/\Delta)\,\phi(\bm b)
.
\end{equation}
This is the saturation scale for a dipole in the fundamental representation. From Eq.~(\ref{eq:xirat}), we have for a dipole in the adjoint representation (as would be appropriate for the gluon distribution),
\begin{equation}
Q_s^2 ( {\bm b},\mbox{adjoint} )=
\frac{C_A}{C_F}\,Q_s^2 ( {\bm b},\mbox{fundamental}  ) = 
\frac{4N_c\pi^2 \alpha_s}{N_c^2 - 1} \,
x G(x,a/\Delta)\,\phi(\bm b)
.
\end{equation}
The name of the model and of the scale $Q_s$ derives from the fact that $\Xi(\bm b, \bm\Delta)$ grows as $\Delta$ increases until it saturates with $\Xi(\bm b, \bm\Delta) \approx 1$ when $\Delta$ reaches approximately $2/Q_s$.

For a specific model, we follow Mueller \cite{mue99} in choosing
\begin{equation}
\label{eq:phimodel}
\phi(\bm b) = \frac{3}{2\pi R_p^3}\, \sqrt{R_p^2 - \bm b^2}\
\theta(|\bm b| < R_p)
.
\end{equation}

\section{Critique of the model}
\label{sec:crit}

The dipole picture and saturation model~\cite{mue99,golec} 
along the lines 
just described has enjoyed some success when its predictions 
are compared to experimental results in both inclusive and 
diffractive deeply inelastic scattering (\cite{golecrev}, and 
references therein). We 
do not attempt a numerical comparison in this paper. However, 
we do offer 
 some comments on the extent to which the dipole picture 
for $f_{q/p}$ should be expected to be reliable.

We have found that the parton distribution function for quarks can be approximated at small $x$ using Eqs.~(\ref{eq:renormalizedfq}) and (\ref{eq:Xiintegral}), 
\begin{equation}
\begin{split}
\label{eq:renormalizedfq2}
xf_{q/p}(x,\mu) ={}&
\frac{N_c}{3 \pi^4}\
\int\!d\bm b
\int\! d{\bm \Delta}\
\theta(\Delta^2\mu^2 > a^2)\
\frac{\Xi(\bm b, {\bm \Delta})}{\Delta^4} 
.
\end{split}
\end{equation}
Clearly,  the model for  $\Xi(\bm b, {\bm \Delta})$ contains 
non-perturbative physics. Furthermore the squared wave function 
$1/\Delta^4$ is a perturbative result that should be 
trusted only for  $\Delta \ll R_p$. Is there any reason to think  
that Eq.~(\ref{eq:renormalizedfq2}) might be reliable at all? 

To examine this issue, first look at the integration range for $\Delta$. There is a renormalization cut $\Delta > a/\mu$ and we may suppose that we consider scale choices such that $a/\mu \ll R_p$. The integration extends to arbitrarily large $\Delta$,  but once $\Delta > 1/Q_s(\bm b)$ we have $\Xi(\bm b,\bm \Delta) \approx 1$ so that the integrand is approximately $1/\Delta^4$. This falloff is sufficiently fast that values of $\Delta$ greater than $1/Q_s(\bm b)$ are not important in the integration. Now, $Q_s(\bm b)$ is proportional to the gluon distribution and at small $x$ there are lots of gluons. For this reason, 
for a central impact parameter $\bm b$, $Q_s(\bm b)$ is larger than 
the normal $300\ {\rm MeV}$ soft hadronic scale.  
With $xG(x) = 10$, $\alpha_s = 0.2$ and $R_p = 4.5\ {\rm GeV}^{-1}$ 
one gets $Q_s(\bm 0, \mbox{fundamental}) \approx 
0.6\ {\rm GeV}$.\footnote{This value is  consistent 
with the value obtained 
by comparison with diffractive DIS data 
in the somewhat different approach~\cite{hs00,diffjet}.}  If 
we were dealing with a large nucleus or with values of $x$ much 
smaller than $10^{-3}$, we could have quite a lot larger values 
of $xG(x)$ and thus a larger saturation 
scale.\footnote{Also, $Q_s(\bm b)$ is larger if we had a 
color $\bm 8$ dipole instead of a color $\bm 3$ dipole, as 
would be the case if we were to investigate the gluon 
distribution.  See \cite{kopel06} for a recent discussion. } Additionally,  $Q_s(\bm b)$  is small  near the  edge of 
the proton. If we were dealing with a large nucleus, the contribution from $\bm b$ near the edge of the nucleus would be less important than for a proton. 

To the extent that $Q_s(\bm b)$ is large,  the 
main contributions to $xf_{q/p}(x,\mu)$ come from regions 
in the integrations in which the model is anchored  in a 
reliable perturbative expansion. But what if $Q_s(\bm b)$ is not so large? Then we must face the facts that 
the model for $\Xi(\bm b, {\bm \Delta})$ is non-perturbative and  
that  the $1/\Delta^4$ squared wave function is a perturbative 
result applied outside the range of validity of the perturbative 
expansion. We can analyze these problems in two ways. 
First, the dipole interaction with the proton, $\Xi(\bm b, {\bm \Delta})$, 
 should be subject to scrutiny.  Second, we can consider 
 what would happen if we were to work at a higher 
 order of perturbation theory. Then we would have 
 new contributions to the partonic state that 
 hits the proton, including the possibility that 
 this state contains more than just two partons.
 
Recall first the  behavior of 
$\Xi(\bm b, {\bm \Delta})$ in the model of 
 Sec.~\ref{sec:satur}, supposing that the 
description of the incoming partonic state as a 
dipole with the perturbative $1/\Delta^4$ squared 
wave function is exactly right. It is indeed true 
that $\Xi(\bm b, {\bm \Delta})$ cannot be reliably 
calculated perturbatively when $\Delta$ is not small. 
However, $\Xi(\bm b, {\bm \Delta})$ corresponds to the 
probability that the dipole scatters. When $\Delta$ is 
large and $|\bm b| < R_p$, it is likely 
that the dipole is almost completely absorbed, which 
 corresponds to $\Xi(\bm b, {\bm \Delta}) \approx 1$. The 
  model  of   Sec.~\ref{sec:satur}  for $\Xi(\bm b, {\bm \Delta})$ 
has this 
 property. Thus $\Xi(\bm b, {\bm \Delta})$ is fixed for 
 small $\Delta$ and for large $\Delta$ as long as $\bm b$ is 
 well inside the proton. It is certainly true 
 that  it is not so well known for 
 intermediate values of $\Delta$ and for large or 
 medium $\Delta$ when $|\bm b| \approx R_p$. 
 In particular, if $Q_s$ is not so  large  one is likely to make 
  an error  by extending the  transparency region 
to intermediate $\Delta$. But the effect is not dramatic, 
  so that  even  here  
  there is not too much that one could do to 
 drastically change $\Xi(\bm b, {\bm \Delta})$ from 
 the form given by the  model. 

Consider now the higher-order states. 
The original eikonal quark plus an antiquark state can 
become an eikonal quark plus an antiquark plus several gluons, 
for example. We could still define a measure $\Delta$ of the 
transverse size of this partonic system. The partonic 
wave function would depend on $\Delta$. It would also 
depend on other dimensionless shape variables that we 
could call $\gamma$. Then we would have a function 
$\Xi_\gamma(\bm b, { \Delta})$, given by 
a matrix element of multi-eikonal 
operators,  describing the 
probability for this state, labeled by $\bm b, {\Delta}$ and 
internal quantum numbers $\gamma$, to 
 scatter. 
This would give an extension of 
Eq.~(\ref{eq:renormalizedfq2}) with the form
\begin{equation}
\label{eq:renormalizedfq3}
xf_{q/p}(x,\mu) =
\int\!d\bm b
\int\! d{\Delta^2}\
\sum_\gamma
|\psi_\gamma(\Delta)|^2\
\Xi_\gamma(\bm b, {\Delta})
.
\end{equation}
The integral needs renormalization, which can 
introduce logarithms of $\Delta\mu$. Except for 
this appearance of the renormalization scale $\mu$, 
the calculation of the wave function  $|\psi_\gamma(\Delta)|^2$ 
involves no hadronic distance scales and no masses. For this 
reason, dimensional analysis tells us that 
$|\psi_\gamma(\Delta)|^2$ is proportional to $\Delta^{-4}$ 
times the logarithms of $\Delta\mu$ times dimensionless 
constants and times factors of $\alpha_s$. This suggests, 
although it certainly does not prove, that the squared wave 
functions $|\psi_\gamma(\Delta)|^2$ are not larger than the 
lowest order result. This leaves us with 
the scattering probabilities $\Xi_\gamma(\bm b, {\Delta})$. We 
do not know the detailed form of these, but it is 
  plausible 
that the complicated states under discussion are 
almost completely absorbed, which corresponds to 
$\Xi_\gamma(\bm b, \Delta) \approx 1$. This is just the 
behavior of the simple dipole version of the scattering probability, 
$\Xi(\bm b, \Delta)$,  for large $\Delta$. 

These arguments do not establish that  Eq.~(\ref{eq:renormalizedfq2}) 
 for the 
quark distribution function must be highly 
accurate if applied to a proton with 
$x \sim 10^{-3}$ rather than, say, a 
very large nucleus or very much lower  $x$. However, 
they do suggest that the picture has enough 
qualitatively right features built into it that it 
should be more useful than would seem from first appearances.

\section{The structure function}
\label{sec:strfun}

The dipole results for structure functions are known from~\cite{mue99}.  
The transverse structure function $F_T$ is given by~\cite{mue99} 
\begin{eqnarray}
F_T 
&=& 
 \frac{1}{4 \pi} \sum_a e_a^2\
\frac{4N_c Q^2}{x(2\pi)^{3}}
\int_0^1\!d\alpha\
[1-2\alpha(1-\alpha)]
\int\! d{\bm b} \int\! d{\bm \Delta}\ 
\nonumber\\
&&\times
\frac{1}{\Delta^2}
\left|
\sqrt{\alpha(1-\alpha)}\,Q  \Delta \
K_0'\left(\sqrt{\alpha(1-\alpha)}\,Q \Delta \right)
\right|^2 
\  
\Xi({\bm b} , {\bm \Delta}
)
, 
\label{FTresult}
\end{eqnarray}
where $Q^2$ is the photon virtuality, and 
$K_0'$ is the derivative of the modified Bessel function. 
The main difference compared to the case of the quark distribution is that 
the ultraviolet region of small $\Delta$ is  now naturally regulated 
by the physical $Q^2$. In Appendix~\ref{app:FT} we 
sketch a derivation of this result along the lines of our derivation 
for the quark distribution function. 

In the remainder of this section, we relate this formula for $F_T$ to the normal factorized form in which $F_T$ is expressed as a sum of perturbatively calculable hard scattering functions $\hat F_T$ convoluted with parton distribution functions. For large $Q^2$, the integral in Eq.~(\ref{FTresult}) is dominated by two integration regions, $\Delta \sim 1/Q$ and $\Delta \gg 1/Q$. We discuss each region in turn.

In the case $\Delta \sim 1/Q \ll R_p$, one can use the small $\Delta$ perturbative formula for $\Xi({\bm b} , {\bm \Delta})$, Eq.~(\ref{eq:cfXigluon}). Then the contribution from this region is a certain one loop integral times $\alpha_s$ times the gluon distribution function. We can recognize that this has the form of a one loop contribution to $\hat F_T$ times the gluon distribution. We do not analyze it further.

The case $\Delta \gg 1/Q$ is more interesting from 
the point of view of this paper. Let us implement the 
requirement $\Delta \gg 1/Q$ in a crude fashion by inserting a 
factor $\theta(Q\Delta > c)$ where $c$ is a fixed number 
of order 1. The only way that we can get a leading contribution to the integral for large $Q\Delta$ without the Bessel function cutting off the integral is for $\alpha(1-\alpha)$ to be small. That is, either $\alpha$ must be small or else $1-\alpha$ must be small. We consider the case $\alpha \ll 1$. To see what this region contributes, we simply neglect $\alpha$ compared to 1 inside the integral, 
\begin{eqnarray}
F_T^{\rm LTq}
&=& 
 \frac{1}{4 \pi} \sum_a e_a^2\
\frac{4N_c Q^2}{x(2\pi)^{3}}
\int_0^\infty\!d\alpha\
\int\! d{\bm b} \int\! d{\bm \Delta}\ 
\theta(Q\Delta > c)
\nonumber\\
&&\times
\frac{1}{\Delta^2}
\left|
\sqrt{\alpha}\,Q \Delta \
K_0'\left(\sqrt{\alpha}\,Q \Delta \right)
\right|^2
\ 
\Xi(
{\bm b} , {\bm \Delta}
)
.
\end{eqnarray}
Here we can change variables from $\alpha$ to $z^2 = \alpha Q^2 \Delta^2$,
giving
\begin{equation}
F_T^{\rm LTq}
= 
\frac{1}{x}
\sum_a e_a^2\
\frac{N_c}{4 \pi^{4}}
\int\! d{\bm b} \int\! d{\bm \Delta}\ 
\theta(Q\Delta > c)\,
\Xi({\bm b} , {\bm \Delta})
\ 
\frac{1}{\Delta^4}
\int_0^\infty\!dz\,z
\left|
z
K_0'\left(z\right)
\right|^2
.
\end{equation}
Using
\begin{equation}
\int_0^\infty\! dz \ z
\left| z K_0'(z)
\right|^2
=\frac{2}{3},
\end{equation}
this is
\begin{eqnarray}
F_T^{\rm LTq}
&=& 
\frac{1}{2x} 
 \sum_a e_a^2\
\frac{N_c}{3\pi^{4}}
\int\! d{\bm b} \int\! d{\bm \Delta}\ 
\frac{\theta(Q\Delta > c)}{\Delta^4}\,
\Xi({\bm b} , {\bm \Delta}
)
.
\label{result1}
\end{eqnarray}
Comparing with Eqs.~(\ref{eq:renormalizedfq}) and (\ref{eq:Xiintegral}),  
we see that we have the lowest order contribution 
to the hard scattering, $\hat F_T$, times the 
dipole form of the quark distribution evaluated 
at a renormalization scale of order $Q/c$. The 
corresponding $1-\alpha \ll 1$ contribution 
gives the same $\hat F_T$ times the antiquark 
distribution. Thus we see that the leading order 
factorization formula works in the 
dipole approximation with the quark 
distribution function defined independently 
according to its definition as the proton 
matrix element of a certain operator.\footnote{Dipole 
contributions 
that are power suppressed with respect 
to the leading factorized term are investigated 
  in \cite{plb06} for the $Q^2$ evolution of the 
  structure function.}

\section{Conclusions} 
\label{sec:conclusions} 

There is an $s$-channel  approximation for 
 structure functions  that is 
quite standard in the literature and is, we believe, well 
motivated. In this approximation, $F_T(x,Q^2)$ is 
given by Eq.~(\ref{FTresult}). This has the form of
a dipole scattering  probability
$\Xi(\bm b, {\bm \Delta})$ convoluted 
with the probability to make the dipole. We have 
presented a variation of the ``saturation''  
model \cite{mue99,golec} for 
$\Xi(\bm b, {\bm \Delta})$ in Eqs.~(\ref{eq:ximodel}), (\ref{eq:Qssq}) 
and (\ref{eq:phimodel}).\footnote{The principle 
refinement is the definition of $x G(x,\mu)$, Eq.~(\ref{eq:Gxdef}).} The 
 approximation (\ref{FTresult}) for $F_T(x,Q^2)$ seems to be 
quite different from the factorized form applicable at large $Q^2$, in 
which $F_T(x,Q^2)$ is expressed as a convolution of a hard 
partonic 
structure function $\hat F_T$  with parton distribution functions. The focus 
of this paper has been to connect these apparently dissimilar pictures 
by investigating the quark distribution function $xf_{q/p}(x,\mu) $ at 
small $x$ using the $s$-channel  picture.

We have found that the parton distribution function for quarks can be approximated at small $x$ using Eqs.~(\ref{eq:renormalizedfq}) and (\ref{eq:Xiintegral}), 
\begin{equation}
\begin{split}
\label{eq:renormalizedfq2bis}
xf_{q/p}(x,\mu) ={}&
\frac{N_c}{3 \pi^4}\
\int\!d\bm b
\int\! d{\bm \Delta}\
\theta(\Delta^2\mu^2 > a^2)\
\frac{\Xi(\bm b, {\bm \Delta})}{\Delta^4} 
.
\end{split}
\end{equation}
This has the form of the same  dipole scattering function $\Xi(\bm b, {\bm
\Delta})$ as in $F_T$, now convoluted with a different  probability to make the
dipole. In fact, the probability to make the dipole is beautifully simple,
\begin{equation}
\frac{N_c}{3 \pi^4}\  \frac{\theta(\Delta^2\mu^2 > a^2)}{\Delta^4}
,
\end{equation}
where $a$ is a calculated number of order 1, Eq.~(\ref{eq:aresult}), 
that accomplishes $\overline{\rm MS}$ renormalization for the quark 
distribution, assuming that $\mu$ is large. 
The power behavior, $1/\Delta^4$, characterizes the squared 
lightcone wave function.

We have seen not only that the quark distribution has a 
simple form in this picture, but  also  that 
the normal lowest order factorized form for $F_T$ 
relates the dipole expression for $F_T$ to the dipole 
expression for $f_{q/p}$. Furthermore, the evolution 
equation for $f_{q/p}$ relates the exponent in 
$\Xi(\bm b, {\bm \Delta})$ to the gluon distribution.

\appendix

\section{Calculation of $u({\bm \Delta})$}
\label{app:little_u}

In this appendix, we compute the integrals for the function $u({\bm \Delta})$ 
introduced in Sec.~\ref{sec:qua}. We begin with Eq.~(\ref{eq:uintegral}),
\begin{equation}
u({\bm \Delta}) =
{2N_c \mu^{4\epsilon} \over (2 \pi)^{6-4\epsilon} } 
\int_0^\infty\! d\Lambda^2  
\int\!d^{2-2\epsilon}{\bm p}_2\int\!d^{2-2\epsilon}{\bm p}_1\
e^{i{\bm \Delta}\cdot({\bm p}_1 - {\bm p}_2)}\
{ 2{\bm p}_1 \cdot {\bm p}_2 \over 
(\Lambda^2 + {\bm p}_2^2 )
(\Lambda^2 + {\bm p}_1^2)} .
\end{equation}
We can introduce two Feynman parameter integrals to put the denominators
into the exponent. This enables us to perform the ${\bm  p}_j$ integrals
\begin{eqnarray}
u({\bm \Delta}) &=&
{4N_c \mu^{4\epsilon} \over (2 \pi)^{6-4\epsilon} } 
\int_0^\infty\! d\Lambda^2  
\int\!d^{2-2\epsilon}{\bm p}_2\int\!d^{2-2\epsilon}{\bm p}_1\
e^{i{\bm \Delta}\cdot({\bm p}_1 - {\bm p}_2)}\
{\bm p}_1 \cdot {\bm p}_2  .
\nonumber\\
&&\times \int_0^\infty \!d\alpha_1\ 
\exp(-\alpha_1(\Lambda^2 + {\bm p}_1^2 ))
\int_0^\infty \!d\alpha_2\ 
\exp(-\alpha_2(\Lambda^2 + {\bm p}_2^2 ))
\nonumber\\
&=&
{4N_c \mu^{4\epsilon} \over (2 \pi)^{6-4\epsilon} } 
\int_0^\infty\! d\Lambda^2  
\int_0^\infty\! d\alpha_1 
\int_0^\infty\! d\alpha_2\ e^{- (\alpha_1 + \alpha_2)\Lambda^2}
\nonumber\\
&&\times \left(-i \frac{\partial}{\partial \Delta_j}\right)
\int\!d^{2-2\epsilon}{\bm p}_1\
\exp(-\alpha_1{\bm p}_1^2  + i{\bm \Delta}\cdot{\bm p}_1 )
\nonumber\\
&&\times \left(i \frac{\partial}{\partial \Delta_j}\right)
\int\!d^{2-2\epsilon}{\bm p}_2\
\exp(-\alpha_2{\bm p}_2^2  - i{\bm \Delta}\cdot{\bm p}_2 )
\nonumber\\
&=&
{4N_c \mu^{4\epsilon} \over (2 \pi)^{6-4\epsilon} } 
\int_0^\infty\! d\Lambda^2  
\int_0^\infty\! d\alpha_1 
\int_0^\infty\! d\alpha_2\ e^{- (\alpha_1 + \alpha_2)\Lambda^2}
\nonumber\\
&&\times \left(-i \frac{\partial}{\partial \Delta_j}\right)
\int\!d^{2-2\epsilon}{\bm p}_1\
\exp(-\alpha_1({\bm p}_1  - i{\bm \Delta}/(2\alpha_1))^2 
-\Delta^2/(4\alpha_1))
\nonumber\\
&&\times \left(i \frac{\partial}{\partial \Delta_j}\right)
\int\!d^{2-2\epsilon}{\bm p}_2\
\exp(-\alpha_2({\bm p}_2  + i{\bm \Delta}/(2\alpha_2))^2 
-\Delta^2/(4\alpha_2))
\nonumber\\
&=&
{4N_c \mu^{4\epsilon} \over (2 \pi)^{6-4\epsilon} } 
\int_0^\infty\! d\Lambda^2  
\int_0^\infty\! d\alpha_1 
\int_0^\infty\! d\alpha_2\ e^{- (\alpha_1 + \alpha_2)\Lambda^2}
\nonumber\\
&&\times \left(i \frac{\Delta_j}{2\alpha_1}\right)
\left(\frac{\pi}{\alpha_1}\right)^{1-\epsilon}
\exp( -\Delta^2/(4\alpha_1))
\nonumber\\
&&\times \left(-i \frac{\Delta_j}{2\alpha_2 }\right)
\left(\frac{\pi}{\alpha_2}\right)^{1-\epsilon}
\exp(-\Delta^2/(4\alpha_2))
\nonumber\\
\nonumber\\
&=&
{N_c \over 2^{6} \pi^{4} } 
\left(4\pi\mu^2\right)^{2\epsilon}\
\Delta^2
\int_0^\infty\! d\Lambda^2  
\int_0^\infty  \frac{d\alpha_1}{\alpha_1}\ \alpha_1^{-1+\epsilon}
\int_0^\infty  \frac{d\alpha_2}{\alpha_2}\ \alpha_2^{-1+\epsilon}
\nonumber\\
&&\times
\exp\left({- (\alpha_1 + \alpha_2)\,\Lambda^2} 
 -\frac{\Delta^2}{4}
\left(\frac{1}{\alpha_1} + \frac{1}{\alpha_2}\right)\right).
\end{eqnarray}

At this point, we can perform the $\Lambda^2$ integral,
\begin{eqnarray}
u({\bm \Delta}) &=&{N_c \over 2^6 \pi^4 } 
\left(4\pi\mu^2\right)^{2\epsilon}\
\Delta^2
\int_0^\infty  \frac{d\alpha_1}{\alpha_1}\ \alpha_1^{-1+\epsilon}
\int_0^\infty  \frac{d\alpha_2}{\alpha_2}\ \alpha_2^{-1+\epsilon}
\nonumber\\
&&\times
\frac{1}{\alpha_1 + \alpha_2}
\exp\left( 
 -\frac{\Delta^2}{4}
\left(\frac{1}{\alpha_1} + \frac{1}{\alpha_2}\right)\right).
\end{eqnarray}
In order to simplify the exponent, we can change variables to $\beta_i =
1/\alpha_i$:
\begin{eqnarray}
u({\bm \Delta}) &=&
{N_c \over 2^6 \pi^4 } 
\left(4\pi\mu^2\right)^{2\epsilon}\
\Delta^2
\int_0^\infty  \frac{d\beta_1}{\beta_1}\ \beta_1^{2-\epsilon}
\int_0^\infty  \frac{d\beta_2}{\beta_2}\ \beta_2^{2-\epsilon}
\nonumber\\
&&\times
\frac{1}{\beta_1 + \beta_2}
\exp\left( 
 -\frac{\Delta^2}{4}
\left(\beta_1 + \beta_2\right)\right)
.
\end{eqnarray}
Now we can change variables to
\begin{eqnarray}
\gamma &=& \beta_1 + \beta_2 , 
\nonumber\\
r &=&\frac{\beta_1}{\beta_1 + \beta_2}.
\end{eqnarray}
The inverse transformation is
\begin{eqnarray}
\beta_1 &=& r\gamma , 
\nonumber\\
\beta_2 &=& (1-r)\gamma.
\end{eqnarray}
The jacobian is
\begin{equation}
\frac{\partial (\beta_1,\beta_2)}{\partial(\gamma,r)} = \gamma.
\end{equation}
Thus
\begin{eqnarray}
u({\bm \Delta}) &=&
{N_c \over 2^6 \pi^4 } 
\left(4\pi\mu^2\right)^{2\epsilon}\
\Delta^2
\int_0^\infty \gamma d\gamma \int_0^1 dr\
 (r\gamma)^{1-\epsilon}\
 ((1-r)\gamma)^{1-\epsilon}
\nonumber\\
&&\times
\frac{1}{\gamma}
\exp\left( 
 -\frac{\Delta^2}{4}\,
\gamma\right)
\nonumber\\
&=&
{N_c \over 2^6 \pi^4 } 
\left(4\pi\mu^2\right)^{2\epsilon}\
\Delta^2
\int_0^\infty  \frac{d\gamma}{\gamma}\ \gamma^{3-2\epsilon}
\exp\left( 
 -\frac{\Delta^2}{4}\,
\gamma\right)
\nonumber\\
&&\times
 \int_0^1 dr\
 (r (1-r))^{1-\epsilon}
.
\end{eqnarray}
We can perform both integrals with the result
\begin{eqnarray}
u({\bm \Delta}) &=&
{N_c \over 2^6 \pi^4 } 
\left(4\pi\mu^2\right)^{2\epsilon}\
\Delta^2
\left(\frac{4}{\Delta^2}\right)^{3-2\epsilon}
\Gamma(3-2\epsilon)\
\frac{\Gamma(2-\epsilon)^2}{\Gamma(4-2\epsilon)}
\nonumber\\
 &=&
{N_c \over 3 \pi^4 }\, \frac{1}{\Delta^4} \
\left(\pi\mu^2 \Delta^2\right)^{2\epsilon}
\frac{\Gamma(2-\epsilon)^2}{1-2\epsilon/3}
.
\end{eqnarray}

\section{An algebraic relation for eikonal operators}
\label{sec:algeb}

In this appendix, we  seek a relation between the operators $ 
{\mbox{Tr}} (  F^\dagger  F  ) $ for the  quark and the gluon
distributions,  where $F$ is given in 
Eq.~(\ref{eikFdef2}). 

Denote by $V$ and $U$ the eikonal operators in the fundamental and
adjoint representation: 
\begin{equation}
\label{VandU}
V ({\bm z}) = F_{\rm{fund.}} ({\bm z}) 
\hspace*{0.2 cm} , \hspace*{0.4 cm}  
U ({\bm z}) = F_{\rm{adj.}} ({\bm z}) \hspace*{0.2 cm} .  
\end{equation}
The following identity holds between $V$ and $U$ at the same point: 
\begin{equation}
\label{VUsamez}
{1 \over 2} U^{a b} ({\bm z}) =  {\mbox{Tr}}  
\left[ t^a V ({\bm z}) t^b V^\dagger ({\bm z}) \right] 
\hspace*{0.2 cm}    , 
\end{equation}
with $t^a$ and $t^b$ generators in the fundamental representation.   This 
can be seen by constructing the adjoint representation  from the product
of $3$ and $\overline 3$. 

Using (\ref{VUsamez}) we can write the trace of two $U$'s at points ${\bm
x}$ and ${\bm y}$ as 
\begin{eqnarray}
\label{Uxy}
{\mbox{Tr}} \left[  U ({\bm x}) U^\dagger ({\bm y}) \right] 
&=&  U^{a b} ({\bm x}) U^{ a  b } ({\bm y})
\nonumber\\
&=& 
4 t^a_{i j} V_{ j l} ({\bm x})  t^b_{l m} V^{\dagger}_{ m i} ({\bm x}) 
t^a_{p q} V_{ q r} ({\bm y})  t^b_{r s} V^{\dagger}_{ s p} ({\bm y})
 \hspace*{0.2 cm} .  
\end{eqnarray}
Now with the identity 
\begin{equation}
\label{fierz}
t^a_{i j} t^a_{p q} = {1 \over 2} \delta_{i q} \delta_{p j} - 
{1 \over {2 N_c} } \delta_{i j} \delta_{p q}
\hspace*{0.2 cm} 
\end{equation}
we obtain 
\begin{eqnarray}
\label{Uxyres}
{\mbox{Tr}} \left[  U ({\bm x}) U^\dagger ({\bm y}) \right] 
&=&  {\mbox{Tr}} \left[  V ({\bm x}) V^\dagger ({\bm y}) \right] 
{\mbox{Tr}} \left[  V^\dagger ({\bm x}) V ({\bm y}) \right] 
\nonumber\\
&-&
 { 1\over N_c} \left\{ 
{\mbox{Tr}} \left[  V ({\bm x}) V^\dagger ({\bm y})  
V ({\bm y}) V^\dagger ({\bm x}) \right] + 
  {\mbox{Tr}} \left[  V ({\bm x}) V^\dagger ({\bm x})  
V ({\bm y}) V^\dagger ({\bm y}) \right]
\right\} 
\nonumber\\
&+& { 1\over N_c^2} {\mbox{Tr}} \left[  V ({\bm x}) V^\dagger ({\bm x}) \right] 
{\mbox{Tr}} \left[  V^\dagger ({\bm y}) V ({\bm y}) \right]
\nonumber\\
&=& 
{\mbox{Tr}} \left[  V ({\bm x}) V^\dagger ({\bm y}) \right] 
{\mbox{Tr}} \left[  V^\dagger ({\bm x}) V ({\bm y}) \right] - 1 
 \hspace*{0.2 cm} . 
\end{eqnarray}

For the operators that appear in the definition of $\Xi$,  
Eq.~(\ref{eq:XiIdef2}), from Eq.~(\ref{Uxyres}) we get
\begin{eqnarray}
\label{oneminus}
{1 \over {N_c^2 - 1}} {\mbox{Tr}} 
\left[ 1 -  U^\dagger ({\bm \Delta}) U ({\bm 0}) \right] &= &
{ C_A \over C_F}  
 {1 \over {N_c}} \ {\mbox{Re}} \  {\mbox{Tr}} 
\left[ 1 -  V^\dagger  ({\bm \Delta}) V({\bm 0}) \right] 
\\
&-& 
{1 \over 2} { C_A \over C_F}  
 {1 \over {N_c^2}} \ | {\mbox{Tr}} 
\left[ 1 -  V^\dagger  ({\bm \Delta}) V({\bm 0}) \right] |^2 
\hspace*{0.2 cm}   . 
\nonumber   
\end{eqnarray}
From this general  relation we recover the simple ratio (\ref{eq:xirat})  in
the case  of small $\bm \Delta$,  where the quadratic term in the right
hand side of  Eq.~(\ref{oneminus}) can be neglected.

\section{The evolution operator for   $F_T$}
\label{app:FT}

In this appendix we derive the $s$-channel formula 
(\ref{FTresult}) 
for the 
structure function $F_T$ in the same fashion  as  was done 
  for the quark distribution in Sec.~\ref{sec:qua}.

We start with the definition of $F_T$,
\begin{equation}
F_T = \frac{1}{8\pi} \int d^4 y\ e^{-iq\cdot y} 
\langle P|J^j(0)J^j(y)|P\rangle .
\end{equation}
Here
\begin{eqnarray}
q &=& \left(
-xP^+,\frac{Q^2}{2xP^+},{\bm 0}
\right),
\nonumber\\
P &=&\left(
P^+,0,{\bm 0}
\right).
\end{eqnarray}
Similarly to Sec.~\ref{sec:qua}, 
we rewrite this as 
\begin{equation}
\label{FTmatrelU}
F_T = 
{\rm Re}\
(2\pi)^{-3}\int\frac{dP^{\prime +}}{2P^{\prime +}}\int\! d{\bm P}'\,
\langle P'|U[A] - U[0]|P\rangle .
\end{equation}
Here $U[A]$ is a function of the field operator $A$,  defined by
\begin{eqnarray}
U[{\cal A}] &=& \frac{P^+}{2\pi}
\int\! d y^+ 
\int\!d{\bm y}_{1}\int\!d{\bm y}_{2}
\int\!dy_1^-\int\!dy_2^-\,
\theta(y_2^- > y_1^-)
\nonumber\\
&&\times
e^{-iq^-y^+} e^{-ixP^+(y_2^- - y_1^-)}
\langle 0|J^j(0,y_2^-,{\bm y}_{2})
J^j(y^+,y_1^-,{\bm y}_{1})|0\rangle_{\cal A} ,
\end{eqnarray}
where the matrix element here is taken 
in an external potential ${\cal A}$.

Now using the interaction picture with 
${\cal A}$ as the perturbation we
have
\begin{eqnarray}
U[{\cal A}] &=& \frac{P^+}{2\pi}
\int\! d y^+ 
\int\!d{\bm y}_{1}\int\!d{\bm y}_{2}
\int\!dy_1^-\int\!dy_2^-\,
\theta(y_2^- > y_1^-)
\nonumber\\
&&\times
e^{-iq^-y^+} e^{-ixP^+(y_2^- - y_1^-)}\
\langle 0|
U(\infty,y_2^-)
J^j(0,y_2^-,{\bm y}_{2})
\nonumber\\
&&\times
U(y_2^-,y_1^-)
J^j(y^+,y_1^-,{\bm y}_{1})
U(y_1^-,-\infty)
|0\rangle_{\cal A} .
\end{eqnarray}
In the approximation that the potential is negligible for large $|y^-|$
while only large positive $y_2^-$ and large negative $y_1^-$ dominate the
integrals, this is
\begin{eqnarray}
U[{\cal A}] &\approx& \frac{P^+}{2\pi}
\int\! d y^+ 
\int\!d{\bm y}_{1}\int\!d{\bm y}_{2}
\int_{-\infty}^0\!dy_1^-\int_0^\infty \!dy_2^-\,
\nonumber\\
&&\times
e^{-iq^-y^+} e^{-ixP^+(y_2^- - y_1^-)}
\nonumber\\
&&\times
\langle 0|
J^j(0,y_2^-,{\bm y}_{2})
U(\infty,-\infty)
J^j(y^+,y_1^-,{\bm y}_{1})
|0\rangle_{\cal A} .
\end{eqnarray}
We understand here that we are going to use the eikonal approximation for
$U$ and if we go beyond the lowest approximation there will be an
effective interval $-y_0^-< y^- < y_0^-$ for $y^-$ inside the
approximation.

We will evaluate this at the lowest order of perturbation theory for the
quantum part of the theory. That is, all of the particles are treated as
free except for the interaction with the external field in $\cal A$. To
carry out this evaluation, we insert intermediate states. The intermediate
states consist of a quark (momentum $k$) and an antiquark (momentum $p$).
These particles carry spin and color, but we choose a notation that
suppresses the spin and color indices. Thus we have
\begin{eqnarray}
U[{\cal A}] &\approx& \frac{P^+}{2\pi}
\int\! d y^+ 
\int\!d{\bm y}_{1}\int\!d{\bm y}_{2}
\int_{-\infty}^0\!dy_1^-\int_0^\infty \!dy_2^-\,
\nonumber\\
&&\times
e^{-iq^-y^+} e^{-ixP^+(y_2^- - y_1^-)}\ (2\pi)^{-12}
\nonumber\\
&&\times
\int_0^\infty\!\frac{dp_2^-}{2p_2^-}\int\! d{\bm p}_2
\int_0^\infty\!\frac{dk_2^-}{2k_2^-}\int\! d{\bm k}_2
\int_0^\infty\!\frac{dp_1^-}{2p_1^-}\int\! d{\bm p}_1
\int_0^\infty\!\frac{dk_1^-}{2k_1^-}\int\! d{\bm k}_1
\nonumber\\
&&\times
\langle 0|
J^j(0,y_2^-,{\bm y}_{2})
|p_2^-,{\bm p}_2,k_2^-,{\bm k}_2\rangle
\nonumber\\
&&\times
\langle p_2^-,{\bm p}_2,k_2^-,{\bm k}_2|
U(\infty,-\infty)
|p_1^-,{\bm p}_1,k_1^-,{\bm k}_1\rangle_{\cal A}
\nonumber\\
&&\times
\langle p_1^-,{\bm p}_1,k_1^-,{\bm k}_1|
J^j(y^+,y_1^-,{\bm y}_{1})
|0\rangle .
\end{eqnarray}
For the matrix element of $U$ we use the (leading) eikonal approximation,
\begin{eqnarray}
\lefteqn{\langle p_2^-,{\bm p}_2,k_2^-,{\bm k}_2|
U(\infty,-\infty)
|p_1^-,{\bm p}_1,k_1^-,{\bm k}_1\rangle_{\cal A}}
\nonumber\\
\quad &=&(2\pi)^2 2p_1^- \delta(p_1^- - p_2^-)
2k_1^- \delta(k_1^- - k_2^-)\
\widetilde F_{\!c}({\bm p}_1 - {\bm p}_2)
\widetilde F({\bm k}_1 - {\bm k}_2).
\end{eqnarray}
Here $F$ is the eikonal factor for the quark, $F_{\!c}$ is the eikonal
factor for the antiquark. In the matrix elements of the current we can use
translation invariance to extract the $y$ dependence. Then
\begin{eqnarray}
U[{\cal A}] &\approx& \frac{P^+}{2\pi}
\int\! d y^+ 
\int\!d{\bm y}_{1}\int\!d{\bm y}_{2}
\int_{-\infty}^0\!dy_1^-\int_0^\infty \!dy_2^-\,
\nonumber\\
&&\times
e^{-iq^-y^+} e^{-ixP^+(y_2^- - y_1^-)}\, (2\pi)^{-10}
\nonumber\\
&&\times
\int_0^\infty\!\frac{dp^-}{2p^-}
\int_0^\infty\!\frac{dk^-}{2k^-}
\int\! d{\bm p}_2
\int\! d{\bm k}_2
\int\! d{\bm p}_1
\int\! d{\bm k}_1
\nonumber\\
&&\times
\langle 0|
J^j(0)
|p^-,{\bm p}_2,k^-,{\bm k}_2\rangle
e^{-i(p_2^+ + k_2^+)y_2^- +i({\bm p}_2 + {\bm k}_2)\cdot {\bm y}_2}
\nonumber\\
&&\times
\widetilde F_{\!c}({\bm p}_1 - {\bm p}_2)
\widetilde F({\bm k}_1 - {\bm k}_2)
\nonumber\\
&&\times
\langle p^-,{\bm p}_1,k^-,{\bm k}_1|
J^j(0)
|0\rangle 
e^{i(p^- + k^-)y_1^+ 
+i(p_1^+ + k_1^+)y_1^-
-i({\bm p}_1 + {\bm k}_1)\cdot {\bm y}_1} .
\end{eqnarray}
Here
\begin{equation}
p_1^+ =\frac{{\bm p}_1^2}{2p^-}, \; 
k_1^+ =\frac{{\bm k}_1^2}{2k^-}, \; 
p_2^+ =\frac{{\bm p}_2^2}{2p^-}, \; 
k_2^+ =\frac{{\bm k}_2^2}{2k^-}.
\end{equation}
We can now perform all of the $y$ integrations to get
\begin{eqnarray}
U[{\cal A}] &\approx& \frac{P^+}{(2\pi)^{6}}
\int_0^\infty\!\frac{dp^-}{2p^-}
\int_0^\infty\!\frac{dk^-}{2k^-}
\int\! d{\bm p}_2
\int\! d{\bm k}_2
\int\! d{\bm p}_1
\int\! d{\bm k}_1
\nonumber\\
&&\times
\langle 0|
J^j(0)
|p^-,{\bm p}_2,k^-,{\bm k}_2\rangle
e^{-i(p_2^+ + k_2^+)y_2^- +i({\bm p}_2 + {\bm k}_2)\cdot {\bm y}_2}
\nonumber\\
&&\times
\widetilde F_{\!c}({\bm p}_1 - {\bm p}_2)
\widetilde F({\bm k}_1 - {\bm k}_2)
\nonumber\\
&&\times
\langle p^-,{\bm p}_1,k^-,{\bm k}_1|
J^j(0)
|0\rangle 
e^{i(p^- + k^-)y_1^+ 
+i(p_1^+ + k_1^+)y_1^-
-i({\bm p}_1 + {\bm k}_1)\cdot {\bm y}_1} 
\nonumber\\
&&\times
\frac{-i}{xP^+ + p_2^+ + k_2^+}\
\frac{-i}{xP^+ + p_1^+ + k_1^+}
\nonumber\\
&&\times
\delta({\bm p}_2 + {\bm k}_2)\,
\delta({\bm p}_1 + {\bm k}_1)\,
\delta(q^- - p^- - k^-) . 
\end{eqnarray}
For the minus-momenta we write $p^- = \alpha q^-$ and $k^- = (1-\alpha)
q^-$. With the use of the delta functions, the momenta are
\begin{eqnarray}
p_1 &=&\left(
\frac{{\bm p}_1^2}{2\alpha q^-},
\alpha q^-,{\bm p}_1
\right),
\nonumber\\
k_1 &=&\left(
\frac{{\bm p}_1^2}{2(1-\alpha)q^-},
(1-\alpha)q^-,-{\bm p}_1
\right),
\nonumber\\
p_2 &=&\left(
\frac{{\bm p}_2^2}{2\alpha q^-},
\alpha q^-,{\bm p}_2
\right) , 
\nonumber\\
k_2 &=&\left(
\frac{{\bm p}_2^2}{2(1-\alpha)q^-},
(1-\alpha)q^-,-{\bm p}_2
\right).
\end{eqnarray}
This gives
\begin{eqnarray}
U[{\cal A}] &\approx& \frac{P^+}{4q^-(2\pi)^{6}}
\int_0^1\!\frac{d\alpha}{\alpha(1-\alpha)}
\int\! d{\bm p}_2
\int\! d{\bm p}_1
\nonumber\\
&&\times
\frac{-i}{xP^+ + p_2^+ + k_2^+}\
\frac{-i}{xP^+ + p_1^+ + k_1^+}\,
{\rm Tr}\{
\widetilde F_{\!c}({\bm p}_1 - {\bm p}_2)
\widetilde F({\bm p}_2 - {\bm p}_1)
\}
\nonumber\\
&&\times
\langle 0|
J^j(0)
|p_2,k_2\rangle
\langle p_1,k_1|
J^j(0)
|0\rangle .
\end{eqnarray}

For the matrix elements of $J^j$, we can write
\begin{eqnarray}
\lefteqn{\langle 0|
J^j(0)
|p_2,k_2\rangle
\langle p_1,k_1|
J^j(0)
|0\rangle
}
\nonumber\\
&=&
\sum_a e_a^2\!\!\sum_{s_1,s_1',s_2,s_2'}
\delta_{s_1s_2}\delta_{s_1's_2'}
\bar u(k_2,s_2)\gamma^j v(p_2,s_2')\
\bar v(p_1,s_1')\gamma^j u(k_1,s_1).\quad\quad
\end{eqnarray}
Now we can insert
\begin{eqnarray}
\delta_{s_1s_2} &=&
\bar u(k_2,s_2)
\gamma^-
u(k_1,s_1)/(2k^-),
\nonumber\\
\delta_{s_1's_2'} &=&
\bar v(p_2,s_2')
\gamma^-
v(p_1,s_1')/(2p^-).
\end{eqnarray}
This leads to
\begin{eqnarray}
\lefteqn{\langle 0|
J^j(0)
|p_2,k_2\rangle
\langle p_1,k_1|
J^j(0)
|0\rangle
}
\nonumber\\
&=&
\frac{1}{4\alpha(1-\alpha)(q^-)^2}
\sum_a e_a^2\,
{\rm Tr}\{
\gamma^j \rlap{/}p_2\,\gamma^-\rlap{/}p_1\,
\gamma^j\rlap{/}k_1\,\gamma^-\rlap{/}k_2
\}
\nonumber\\
&=&
4 \sum_a e_a^2\
\frac{1-2\alpha(1-\alpha)}
{\alpha(1-\alpha)}
\,{\bm p}_1\cdot {\bm p}_2 .
\end{eqnarray}
Thus
\begin{eqnarray}
U[{\cal A}] &\approx& 
\frac{1}{4 \pi} \sum_a e_a^2\
\frac{2P^+}{q^-(2\pi)^{5}}
\int_0^1\!d\alpha
\int\! d{\bm p}_2
\int\! d{\bm p}_1\
\frac{1-2\alpha(1-\alpha)}
{\alpha^2(1-\alpha)^2}
\,{\bm p}_1\cdot {\bm p}_2
\nonumber\\
&&\times
\frac{-i}{xP^+ + p_2^+ + k_2^+}\
\frac{-i}{xP^+ + p_1^+ + k_1^+}
\nonumber\\
&&\times
{\rm Tr}\{
\widetilde F_{\!c}({\bm p}_1 - {\bm p}_2)
\widetilde F({\bm p}_2 - {\bm p}_1)
\} .
\end{eqnarray}

We can rewrite the energy denominators to obtain
\begin{eqnarray}
U[{\cal A}] &\approx& 
- \frac{1}{4 \pi} \sum_a e_a^2\
\frac{4Q^2}{x(2\pi)^{5}}
\int_0^1\!d\alpha
\int\! d{\bm p}_2
\int\! d{\bm p}_1\
[1-2\alpha(1-\alpha)]
\,{\bm p}_1\cdot {\bm p}_2
\nonumber\\
&&\times
\frac{1}{\alpha(1-\alpha)Q^2 + {\bm p}_2^2}\
\frac{1}{\alpha(1-\alpha)Q^2 + {\bm p}_1^2}\
\nonumber\\
&&\times
{\rm Tr}\{
\widetilde F_{\!c}({\bm p}_1 - {\bm p}_2)
\widetilde F({\bm p}_2 - {\bm p}_1)
\} .
\end{eqnarray}

With 
\begin{eqnarray}
\widetilde F_{\!c}({\bm p}_1 - {\bm p}_2)
\widetilde F({\bm p}_2 - {\bm p}_1) &=&
\int d{\bm b}\,d{\bm \Delta}\
e^{i({\bm p}_1 - {\bm p}_2)\cdot({\bm b} + {\bm \Delta}/2)}
e^{i({\bm p}_2 - {\bm p}_1)\cdot({\bm b}- {\bm \Delta}/2) }
\nonumber\\
&& \times
F_{\!c}({\bm b} + {\bm \Delta}/2)\,
F({\bm b} - {\bm \Delta}/2) , 
\end{eqnarray}
we get 
\begin{eqnarray}
\label{finalU}
U[{\cal A}] &\approx& 
- \frac{1}{4 \pi} \sum_a e_a^2\
\frac{4Q^2}{x(2\pi)^{5}}
\int_0^1\!d\alpha\
[1-2\alpha(1-\alpha)]
\int\! d{\bm b} \int\! d{\bm \Delta}\ 
\nonumber\\
&&\times
{\rm Tr}\{
F_{\!c}({\bm b} + {\bm \Delta}/2)\,
F({\bm b} - {\bm \Delta}/2 )
\} 
\nonumber\\
&&\times
\int\! d{\bm p}_2\
e^{-i {\bm p}_2\cdot {\bm \Delta}}
\frac{{p}_2^j}{\alpha(1-\alpha)Q^2 + {\bm p}_2^2}\
\nonumber\\
&&\times
\int\! d{\bm p}_1\
e^{i {\bm p}_1\cdot {\bm \Delta}}
\frac{{p}_1^j}{\alpha(1-\alpha)Q^2 + {\bm p}_1^2}\
.
\end{eqnarray}
We now take the hadron matrix element (\ref{FTmatrelU}), and use  
the definition (\ref{eq:XibDeltadef}). Then   
performing the integrations over ${\bm p}_1$ and ${\bm p}_2$  gives  
Eq.~(\ref{FTresult}).



\begin{references}


\bibitem{jjj}
J.M.~Campbell, J.W.~Huston  and  W.J.~Stirling, 
Rept.\ Prog.\ Phys.\ {\bf 70} (2007) 89. 

\bibitem{ericemue}
    A.H.~Mueller, hep-ph/0501012, in Proc.
    ``{\em QCD at cosmic energies}" (Erice 2004); 
   E.~Iancu,  A.H.~Mueller and S.~Munier, 
   Phys.\ Lett.\ {\bf B}606  (2005) 342;  
   A.H.~Mueller, A.I.~Shoshi and S.M.H.~Wong, 
   Nucl.\ Phys.\ {\bf B715} (2005) 440; 
   Y.~Hatta, E.~Iancu, L.~McLerran, A.~Stasto and 
   D.N.~Triantafyllopoulos, Nucl.\ Phys.\ {\bf A764} 
   (2006) 423.   

\bibitem{golecrev} 
    K.~Golec-Biernat,     
    Acta Phys.\ Polon.\  B{\bf 35}  (2004) 3103; 
    hep-ph/0507251, in Proc. ``{\em Baryons 2004}",  
    Nucl.\ Phys.\ {\bf A755}  (2005) 133; 
    K.~Golec-Biernat and  M.~W{\" u}sthoff,
    Phys.\ Rev.\ D {\bf 60}  (1999) 114023. 

\bibitem{mue99}
    A.H.~Mueller,    hep-ph/0111244, lectures at  Cargese
    Summer School;   Nucl.\  Phys.\ {\bf B558} (1999) 285.  
    

\bibitem{cs82}
J.C.~Collins and D.E.~Soper, Nucl.\  Phys.\ {\bf B194} (1982)  445. 

\bibitem{plb06}
F.~Hautmann,   Phys.\ Lett.\ {\bf B}643  (2006) 171.


\bibitem{hks} 
F.~Hautmann, Z.~Kunszt and D.E.~Soper, 
Nucl.\  Phys.\ {\bf B563} (1999) 153;   
    Phys.\ Rev.\ Lett.\ {\bf 81} (1998) 3333; see also 
J.D.~Bjorken, J.~Kogut and D.E.~Soper, 
    Phys.\ Rev.\  D {\bf 3} (1971) 1382. 



\bibitem{hs00} 
F.~Hautmann and D.E.~Soper, Phys.\ Rev.\  D {\bf 63} (2000) 011501.  

  
\bibitem{golec}
    J.~Bartels, K.~Golec-Biernat and H.~Kowalski,
    Phys.\ Rev.\ D {\bf 66} (2002) 014001.

\bibitem{diffjet}
    F.~Hautmann, JHEP {\bf 0210} (2002) 025, 
    JHEP {\bf 0204} (2002) 036.  

\bibitem{kopel06}
    B.Z.~Kopeliovich, B.~Povh and I.~Schmidt, hep-ph/0607337. 

\end{references}
\end{document}